\documentclass[12pt]{article}
\usepackage[top=3cm, left=3cm, right=3cm, bottom=3cm]{geometry}
\usepackage{floatrow}
\usepackage[utf8]{inputenc}
\usepackage[english]{babel}
\usepackage{graphicx}
\usepackage[T1]{fontenc}
\usepackage{amsmath, amssymb}
\usepackage{natbib}
\usepackage{float}
\floatstyle{plaintop}
\restylefloat{table}
\usepackage{multirow}
\usepackage{parskip}
\usepackage{times}
\usepackage[font=small,labelfont=bf]{caption}
\usepackage{subcaption}
 \usepackage[bottom]{footmisc}
\usepackage{hyperref}
\usepackage{multicol}
\usepackage{lipsum}
\usepackage{authblk}
\usepackage{dcolumn}
\usepackage{animate}
\usepackage{multimedia}
\usepackage{media9}
\usepackage[Export]{adjustbox}
\usepackage{tabularx}
\usepackage{cellspace}
\usepackage{longtable}
\usepackage{booktabs}
\renewcommand\tabularxcolumn[1]{>{\centering\arraybackslash}S{p{#1}}}

\DeclareCaptionLabelSeparator{pipe}{ $\textbf{|}$ }

\let\oldnl\nl
\newcommand{\nonl}{\renewcommand{\nl}{\let\nl\oldnl}}

\usepackage[title]{appendix}
\usepackage{algorithm2e}

\DeclareMathOperator*{\argmin}{arg\,min}
\usepackage{caption}

\usepackage{algorithmic}

\makeatletter
\renewcommand{\@algocf@capt@plain}{above}
\makeatother

 \providecommand{\keywords}[1]
{
  \small	
  \textbf{\textit{Keywords---}} #1
}

\title{Learning Individual Reproductive Behavior from Aggregate Fertility Rates via Neural Posterior Estimation}

\author[1,2]{Daniel Ciganda}
\author[1,2]{Ignacio Campón}
\author[3,4]{Iñaki Permanyer}
\author[5,6]{Jakob H Macke} 

\affil[1]{\small{Max Planck Institute for Demographic Research}}
\affil[2]{\small{Statistics Institute, UDELAR}}
\affil[3]{\small{Center For Demographic Studies, Autonomous University of Barcelona}}
\affil[4]{\small{ICREA, Catalan Institution for Research and Advanced Studies}}
\affil[5]{\small{Machine Learning in Science, University of Tübingen \& Tübingen AI Center}}
\affil[6]{\small{Department Empirical Inference, Max Planck Institute for Intelligent Systems}}

\date{}

\begin{document}
\setlength{\parindent}{0pt}

\maketitle              
\begin{abstract}
Age-specific fertility rates (ASFRs) provide the most extensive record of reproductive change, but their aggregate nature obscures the individual-level behavioral mechanisms that drive fertility trends. To bridge this micro-macro divide, we introduce a likelihood-free Bayesian framework that couples a demographically interpretable, individual-level simulation model of the reproductive process with Sequential Neural Posterior Estimation (SNPE). We show that this framework successfully recovers core behavioral parameters governing contemporary fertility, including preferences for family size, reproductive timing, and contraceptive failure, using only ASFRs. The framework's effectiveness is validated on cohorts from four countries with diverse fertility regimes. Most compellingly, the model, estimated solely on aggregate data, successfully predicts out-of-sample distributions of individual-level outcomes, including age at first sex, desired family size, and birth intervals. Because our framework yields complete synthetic life histories, it significantly reduces the data requirements for building microsimulation models and enables behaviorally explicit demographic forecasts.
\end{abstract}

\vfill \keywords{simulation-based inference, individual-level modelling, age-specific fertility rates, reproductive behavior, microsimulation}

\section{Introduction} 
\label{intro}

Globally declining fertility rates are driving a profound demographic transformation, presenting first-order societal challenges and underscoring the need for robust models to understand and anticipate future fertility trends. Age-specific fertility rates (ASFRs), defined as the number of births to women of a given age per person-year of exposure, remain the workhorse input for these tasks because the underlying birth counts are collected continuously in every country’s vital-statistics system. ASFRs series therefore provide unparalleled temporal and geographic coverage.

Yet ASFRs are only marginal aggregates. They reveal \emph{when} births occur but not \emph{why}. A long tradition of parametric models has been developed to describe ASFRs curves, smooth observed schedules, extrapolate fitted parameters, or convert projected total fertility rates into age patterns for cohort-component population forecasts \citep{Coale1974, brass1974perspectives, schmertmann2003system, chandola1999recent}. While existing approaches can excel at capturing aggregate ASFRs shapes, their parameters are typically macro-level abstractions. As such, they offer limited insight into the individual behavioural mechanisms driving fertility decisions \citep{hoem1989impact}. Consequently, forecasts based on these models often cannot explicitly account for how evolving individual choices regarding the timing of parenthood, contraceptive practice, and birth-spacing and stopping decisions shape future fertility trends.

This disconnect is striking in light of the explosion of micro‑level fertility research based on rich individual data sources. Studies using birth histories, time-use diaries, and linked administrative records have uncovered nuanced relationships between education, labor market trajectories, union stability, and childbearing. Yet virtually none of that behavioural insight is incorporated into fertility projection models. The challenge of understanding contemporary fertility lies in bridging this micro-macro divide.

This paper proposes a step toward this goal. We introduce a demographically interpretable individual-level model that captures key features of modern reproductive behavior, including stopping, spacing, and contraceptive failure. Estimating such a richly parameterised model from coarse data is a significant challenge, as many combinations of micro-parameters can generate similar ASFRs curves. We overcome this challenge by pairing our individual-level, mechanistic model with Sequential Neural Posterior Estimation (SNPE), a powerful simulation-based inference method. This approach allows us to demonstrate that core behavioral parameters governing the overall level and timing of fertility can be recovered from ASFRs alone. We then fit the model to U.S. National Survey of Family Growth cohorts and to Demographic and Health Survey cohorts from Colombia, the Dominican Republic, and Peru, representing four settings that span markedly different fertility regimes.

The framework's effectiveness is confirmed through rigorous validation. First, posterior predictive checks show that the observed aggregate fertility rates are highly plausible under the model, falling well within the 95\% posterior predictive intervals. Most significantly, we conduct out-of-sample validation, demonstrating that the model can successfully predict distributions of individual-level characteristics, such as the empirical distributions of age at first sexual intercourse, desired family size, and birth intervals, none of which inform the estimation step. 

By establishing a demographically interpretable and statistically robust link between individual reproductive behaviors and population-level fertility outcomes, our framework offers a promising path towards a more nuanced understanding of fertility dynamics. This approach not only enables more accurate, behaviorally grounded population forecasts but also lowers the barrier to building detailed microsimulation models by significantly reducing their data requirements.

\subsection{Previous Work}

Efforts to model the reproductive process mechanistically have a long and distinguished tradition in demography. Pioneered by early work on fecundability \citep{gini1924premieres} and significantly advanced by work on 'natural fertility' \citep{Henry1953}, this foundation led to a first wave of mathematical and early simulation models of the reproductive process \citep[among others]{brass1958distribution, singh1963probability, ridley1966analytic, potter1972additional, bongaarts1977dynamic}. These models sought to capture the biological and behavioral drivers of family building with increasing realism.

However, this vibrant tradition of individual-level modeling did not continue to evolve within demography at the same pace or in the same manner as in other scientific fields. While empirical fertility research using micro-data flourished \citep{xie2000demography}, the further development of these process-based simulators saw less sustained momentum. More critically, this stream of work remained largely disconnected from the concurrent revolution in statistical methodology for simulation-based inference. Techniques like Approximate Bayesian Computation (ABC) or other simulation-based inference (SBI) approaches that became influential in fields like ecology and genetics \citep{beaumont2010approximate, hartig2011statistical}, were not systematically integrated in demography. This methodological lag created a persistent gap in formally linking mechanistic understanding to empirical observations in fertility research.

Our own previous work showed that such a bridge is feasible in natural-fertility contexts (historical or religious communities without deliberate birth control) where a handful of biological parameters govern the reproductive process and can be inferred from ASFRs alone \citep{ciganda2024modelling}. The present paper tackles the far more ambitious and policy-relevant task of modeling contemporary fertility, where stopping, spacing, and imperfect contraception introduce complex decision-making and heterogeneity in reproductive dynamics.

\subsection{Organization of the Article}

The remainder of this paper is structured as follows: Section \ref{sec:model} describes our individual-level computational model of the reproductive process. Section \ref{data} details the data sources and the construction of the main demographic quantities used in our analysis. Section \ref{sec:methods} outlines the SNPE methodology. The Results section presents our findings on parameter identifiability via cross-validation, model fit through posterior predictive checks, and the out-of-sample validation against micro-level data. Finally, we discuss the implications of our findings, limitations, and avenues for future research in the Discussion.

\section{Model}
\label{sec:model}

The transition from a natural to a modern fertility regime in human populations is marked by the emergence of \textit{stopping} and \textit{spacing} behaviors (deciding whether and when to have children), representing a shift from a reproductive process largely determined by physiological factors to one shaped by individual preferences. As a result, fertility patterns become more diverse, with family formation increasingly tailored to individual or couple-level choices rather than biological constraints alone.

Building on these insights, our model represents the dynamics of reproductive behavior in contemporary populations by placing fertility intentions and desires at its core. Specifically, it conceptualizes childbearing decisions as the interplay between physiological capacity, access to contraception, and individuals’ subjective preferences, such as desired family size and timing.

We employ an individual-level, discrete-time simulation model. A cohort of $N$ women is simulated from the age of potential sexual activity until the end of their reproductive capacity, with their state updated in monthly time steps.

The model begins by assigning each woman key characteristics that will shape her reproductive life. These include two critical ages: the \emph{age at sexual initiation}, when she first becomes exposed to the risk of childbearing, and the \emph{age at intentional reproduction}, representing the point at which she actively starts trying to conceive. Because these two ages often differ, there is typically a period during which a woman is sexually active but not yet intending to have children, an interval associated with the risk of unplanned pregnancies.

Additionally, each woman is assigned a \emph{desired family size}, reflecting the number of children she ultimately hopes to have, and a \emph{desired number of months between births}, which influences her birth spacing behavior.

In each monthly step, a woman’s probability of conception is evaluated. This probability depends on her current age, her current parity relative to her desired family size, whether she has reached her age at intentional reproduction, and, if she has had prior births, whether her desired birth spacing interval has elapsed.

Specifically, when a woman is past her age at intentional reproduction, has not yet reached her desired family size, and is past her desired birth spacing interval (if applicable), we assume she is actively trying to conceive and uses no contraception. Conversely, if she has not yet reached her age at intentional reproduction, or has already achieved her desired family size, or is within her desired spacing interval, she is assumed to be using contraception. This results in a reduced, but non-zero, probability of conception, accounting for imperfect contraceptive effectiveness and the possibility of unplanned pregnancies throughout the reproductive lifespan.

Whenever a conception occurs, the woman is considered non-susceptible for nine months of pregnancy, followed by an additional three-month period of postpartum amenorrhea, during which her probability of another conception is zero. After giving birth, a further delay is introduced to represent the woman’s desired number of months between births, mirroring real-life behaviors around birth spacing.

\subsection{Implementation}

The dynamics described above are implemented by assigning the following characteristics to each woman $i$ and defining the processes that govern her reproductive trajectory:

\begin{enumerate}
\item \textbf{Age at Sexual Initiation ($X_i$):} Each woman's age at sexual initiation, $X_i$, is drawn from a lognormal distribution. This distribution is parameterized by a mean age $\mu_s$ and a standard deviation $\sigma_s$ (both in years), which are converted to the distribution's underlying parameters (the mean of logarithms, $\mu_{\ln,s}$, and the standard deviation of logarithms, $\sigma_{\ln,s}$) using the standard formulas: $\mu_{\ln,s} = \ln(\mu_s^2 / \sqrt{\mu_s^2 + \sigma_s^2})$ and $\sigma_{\ln,s} = \sqrt{\ln(1 + \sigma_s^2 / \mu_s^2)}$. The value for each woman is drawn in years and then converted to months.

    \item \textbf{Age at Intentional Reproduction ($R_i$):} $R_i$ (in months) is also drawn from a lognormal distribution. The mean of this distribution is set to $\mu_r = \mu_s + \delta_r$, where $\delta_r$ (in years) is an estimated model parameter. The parameters $\mu_r$ and $\sigma_r$ (in years) are used to derive the underlying lognormal parameters $\mu_{\ln,r}$ and $\sigma_{\ln,r}$ using the same conversion formulas.

    \item \textbf{Desired Family Size ($D_i$):} $D_i$ is drawn from a Weibull distribution, parameterized by its mean $\mu_d$ and standard deviation $\sigma_d$, which are used to approximate the Weibull shape parameter $\alpha_d = (\sigma_d / \mu_d)^{-1.086}$ and scale parameter $\lambda_d = \mu_d / \Gamma(1 + 1/\alpha_d)$. Samples are rounded to the nearest integer.

\item \textbf{Desired Birth Spacing ($B_i$):} $B_i$ (in months) is drawn from a lognormal distribution with a mean desired spacing $\mu_b$ and a standard deviation $\sigma_b$ (both in months). These parameters are used to derive the underlying mean and standard deviation for the lognormal distribution.
\end{enumerate}

The age-specific monthly probability of conception (fecundability), $\phi(x)$ at age $x$ (in years), is a crucial component of the model. It is well established that fecundability typically increases from puberty, reaches a peak during the early adult years, and progressively declines thereafter towards menopause \citep{bendel1978estimate,Larsen2000,weinstein1990components}. To capture this characteristic age pattern, we model $\phi(x)$ over a reproductive window from $x_{min}=10$ to $x_{max}=50$ years. Age $x$ is first rescaled to $x_s = (x - x_{min}) / (x_{max} - x_{min})$, so $x_s \in [0,1]$. Instead of using a standard polynomial basis, we employ Bernstein basis polynomials, which provide flexibility and ensure the curve is well-behaved at the boundaries of the reproductive window. Specifically, $\phi(x)$ is modeled as a linear combination of two Bernstein basis polynomials of degree 3:
\[
\phi(x) = \beta_1 [3x_s(1-x_s)^2] + \beta_2 [3x_s^2(1-x_s)]
\]
The terms $3x_s(1-x_s)^2$ and $3x_s^2(1-x_s)$ correspond to the two intermediate basis functions ($B_{1,3}(x_s)$ and $B_{2,3}(x_s)$) of a Bernstein basis of degree 3. The exclusion of the first and last basis functions from the model inherently forces the curve $\phi(x)$ to be zero at the start ($x_s=0$) and end ($x_s=1$) of the reproductive window. The parameters $\beta_1$ and $\beta_2$ are estimated and control the level and shape of the fecundability curve, allowing for a variety of empirically plausible age-fecundability profiles.

The actual probability of conception in a given month depends not only on this baseline age-specific fecundability but also on a woman's reproductive intentions and contraceptive use. As outlined earlier, if a woman $i$ with current parity $k_i$ is \emph{actively trying to conceive} (i.e., her current age is $\ge R_i$, her parity $k_i < D_i$, and sufficient time $B_i$ has elapsed since her last birth), her probability of conception is $\phi(x)$.

Conversely, if a woman $i$ with current parity $k_i$ is sexually active but \emph{not} actively trying to conceive, she is assumed to use contraception. Her monthly probability of conception is then $\kappa_{eff} \phi(x)$. The parameter $\kappa \in (0,1)$ is an estimated baseline probability of contraceptive failure. To model an \emph{increased contraceptive vigilance once desired parity ($D_i$) is met}, the effective failure rate $\kappa_{eff}$ is defined as $\kappa$ if her current parity $k_i < D_i$, and is reduced to $\kappa^2$ if $k_i \ge D_i$. This formulation reflects intensified efforts to avoid further births, substantially reducing the risk of an unplanned pregnancy (assuming $\kappa < 1$) after a woman reaches her desired family size.

Table~\ref{tab:model_parameters} summarizes the estimated model parameters.

\begin{table}[H]
\centering
\begin{tabular}{lp{13cm}}
\toprule
\textbf{Parameter} & \textbf{Description} \\
\midrule
$\mu_s$   & Mean age at sexual initiation (in years), for a lognormal distribution. \\
$\sigma_s$ & Standard deviation of the age at sexual initiation. \\
$\delta_r$ & Mean gap (in years) between sexual initiation and intentional reproduction. \\
$\sigma_r$ & Standard deviation of the age at start of intentional reproduction. \\
$\mu_d$   & Mean desired family size, for a Weibull distribution. \\
$\sigma_d$ & Standard deviation of desired family size. \\
$\mu_b$   & Mean desired birth spacing (in months), for a lognormal distribution. \\
$\sigma_b$ & Standard deviation of the desired birth spacing. \\
$\kappa$  & Probability of contraceptive failure per month. \\
$\beta_1$ & Defines the peak of the age-specific fecundability curve. \\
$\beta_2$ & Defines the decline of the age-specific fecundability curve. \\
\bottomrule
\end{tabular}
\caption{Model parameters and their descriptions.}
\label{tab:model_parameters}
\end{table}

\section{Data}
\label{data}

Our analysis draws on two large, nationally representative sources. For countries outside the United States, we use the Demographic and Health Surveys (DHS), a leading initiative of the United States Agency for International Development (USAID) that has conducted over 300 surveys in more than 90 low-and middle-income countries. Each DHS round collects complete retrospective birth histories, allowing the reconstruction of every woman’s fertility trajectory. Data for the United States are drawn from several early cycles of the National Survey of Family Growth (NSFG). These surveys, conducted by the National Center for Health Statistics (NCHS), are designed to provide detailed national data on family life, marriage, divorce, contraception, and importantly for this study, complete pregnancy and birth histories. We utilize these data as compiled and harmonized by the Integrated Fertility Survey Series (IFSS), applying IFSS sampling weights to generate population-level estimates. For both DHS and NSFG data, to achieve adequate sample sizes while ensuring cohorts of women were exposed to broadly similar social conditions, we pool available survey rounds and organize respondents into 10-year birth cohorts. The final cohorts analyzed correspond to women born between 1938–1948 for the United States, and between 1966–1975 for Colombia, the Dominican Republic, and Peru.

A key consideration in selecting empirical settings was to align them with the behavioral logic embedded in our model. Thus, we narrowed the DHS universe to countries where adolescent childbearing is primarily unintended. This was operationalized by calculating, for each potential cohort, the proportion of births to women under 18 that were declared as unplanned, retaining only those populations where this share exceeded one-half. This selection criterion identifies populations where, for a significant fraction of teenagers, sexual activity and intentional reproduction are decoupled—a precondition for our model's assumption that early fertility is largely viewed as undesirable. It also serves to exclude contexts where prevailing norms, such as marriage-centered childbearing from a young age, would imply a data-generating process markedly different from the one our model specifies.

Because DHS sample sizes and the number of survey rounds differ widely across countries, we implemented a further filter based on cohort size. We selected DHS countries where the chosen cohort included a minimum of 4,000 women, a threshold established to guarantee reasonably precise estimates of both age-specific total and unplanned fertility rates across the entire reproductive span. This resulted in the inclusion of three DHS countries: Colombia, Peru, and the Dominican Republic. Together with the United States, these four nations constitute our study sample, offering a valuable cross-section of different fertility patterns and socio-economic contexts that allows for a robust examination of our model's performance across diverse demographic and social settings.

We compute the conventional age-specific fertility rate by dividing the number of births at each age by the number of person-years of exposure contributed at that age. To test how information on birth intentionality improves our estimates, we also construct an age-specific unplanned fertility rate (ASUFRs). This rate is defined as the number of \textit{unplanned} births at a given age divided by person-years of exposure. The classification of a birth as unplanned is based on a harmonized strategy across the DHS and NSFG that combines direct survey questions about birth timing with a parity check (i.e., whether the birth exceeded the mother's ideal family size). This hybrid approach is designed to mitigate recall bias and ex-post rationalization, which are known to affect estimates based solely on intention questions \citep{casterline2007estimation}.

\section{Simulation-Based Inference}
\label{sec:methods}

Inferring parameters for complex process-based models is often challenging due to intractable likelihood functions. Neural Posterior Estimation (NPE) offers a powerful simulation-based inference  approach that circumvents these difficulties by directly learning an approximation to the posterior distribution \(p(\theta \mid x)\) from simulated data pairs \((\theta, x)\). In practice, NPE proceeds by drawing parameters \(\theta\) from a prior distribution \({p}(\theta)\), using the simulator to generate the corresponding data \(x \sim p(x \mid \theta)\), and then training a flexible neural density estimator, \(q_\phi(\theta \mid x)\) (with network parameters \(\phi\)), on these simulated pairs to approximate the true posterior.

While basic NPE is effective, our study employs a powerful sequential variant known as Automatic Posterior Transformation (APT), sometimes referred to as SNPE-C \citep{greenberg2019automatic}. Like other sequential neural posterior estimation methods, APT enhances inferential efficiency by drawing parameter samples from an iteratively updated proposal distribution, $\tilde{p}(\theta)$, rather than solely from the fixed prior. The key challenge in this approach is to correct for the use of a proposal distribution, as naively training on the resulting samples would lead to an incorrect posterior estimate. APT solves this by analytically transforming the posterior estimate itself. For each simulated data point drawn from a proposal $\tilde{p}_r(\theta)$, it re-weights the target density using the known ratio $\tilde{p}_r(\theta)/p(\theta)$. The network is then trained by minimizing the loss with respect to this transformed "proposal posterior" (see line 10 of Algorithm \ref{alg:apt}). A key advantage of this formulation is that because the loss is valid for any proposal, the network can be trained on the cumulative set of all simulations generated across all previous rounds. This procedure allows the network to learn the true posterior without relying on the potentially unstable importance weights used by other methods.

A key strength of the APT framework is its flexibility, as it supports a wide range of density estimators, including mixture-density networks and various types of normalizing flows. In this study, we employ a Neural Spline Flow (NSF), an expressive and powerful density estimator based on monotonic rational-quadratic splines \citep{durkan2019neural}.

This adaptive refinement progressively concentrates simulation effort on regions of high posterior probability, leading to a more accurate posterior approximation with fewer overall simulations. SNPE and related neural posterior estimation techniques have found successful application in diverse scientific domains, including neuroscience \citep{gonccalves2020training, groschner2022biophysical, deistler2022energy}, astrophysics \citep{dax2025real}, cognitive science \citep{von2022mental}, and exoplanet searches \citep{vasist2023neural}, among others.

Algorithm~\ref{alg:apt} summarizes the complete APT procedure employed in this study, which was implemented using the \texttt{sbi} Python package \citep{boelts2024sbi}.

\begin{algorithm}[H]
\SetAlgoLined
\caption{Automatic Posterior Transformation (APT)}
\label{alg:apt}

\KwIn{Simulator with implicit likelihood $p(x|\theta)$, observed data $x_{o}$, prior $p(\theta)$, conditional density family $q_{F(x,\phi)}(\theta)$, number of rounds $R$, simulations per round $N$.}
\KwOut{The final posterior approximation $q_{F(x_{o},\phi)}(\theta)$.}
\BlankLine

Set initial proposal distribution $\tilde{p}_{1}(\theta) \leftarrow p(\theta)$\;

\For{$r=1$ \KwTo $R$}{

\For{$j=1$ \KwTo $N$}{
    Sample $\theta_{r,j} \sim \tilde{p}_{r}(\theta)$\;
    Simulate $x_{r,j} \sim p(x|\theta_{r,j})$\;
}

Update network weights $\phi$ by minimizing the loss over all collected data $\{(\theta_{i,j}, x_{i,j})\}$ from rounds $i=1...r$:

$\phi \leftarrow \argmin_{\phi} \sum_{i=1}^{r} \sum_{j=1}^{N} -\log \tilde{q}_{x_{i,j},\phi}(\theta_{i,j})$\;

where $\tilde{q}_{x,\phi}(\theta) := q_{F(x,\phi)}(\theta) \frac{\tilde{p}(\theta)}{p(\theta)} \frac{1 }{Z(x,\phi)}$ with $Z(x,\phi) = \int_\theta q_{F(x,\phi)}(\theta) \frac{\tilde{p}(\theta)}{p(\theta)} d\theta$\;

Set the next proposal to the current posterior estimate for the observed data $x_o$:

$\tilde{p}_{r+1}(\theta) \leftarrow q_{F(x_{o},\phi)}(\theta)$\;
}
\end{algorithm}

The code and data used in this study, along with detailed instructions for replication, can be found on GitHub at \href{https://github.com/dciganda/comfert_regulated}{github.com/dciganda/comfert\_regulated}. The repository is currently private but access can be granted upon request to the corresponding author.

\subsection{Inference Scenarios}

Simulation-Based Inference typically relies on summarizing potentially high-dimensional data into lower-dimensional, sufficiently informative statistics. This can be done either through explicit, pre-defined summaries or via learned embeddings. The ability to work with a low-dimensional yet descriptive representation of the data is often essential for making parameter inference tractable, especially for complex simulators.

When working with aggregate data, as in our study, the dimensionality reduction is an inherent feature of the observation process. For instance, vital statistics or administrative registers implicitly aggregate thousands of individual fertility trajectories into relatively low-dimensional summaries like age-specific fertility rates. We thus inherit a pre-summarized dataset, limiting our ability to engineer or learn optimal summary statistics directly from the underlying micro-level process.

A critical question then arises: do these aggregate data provide sufficient information to tightly constrain the posterior distributions of the parameters in our individual-level model? The core challenge is that different combinations of micro-parameters (e.g., those governing desired family size, timing of intentional reproduction, or contraceptive failure) could potentially yield very similar aggregate fertility schedules, resulting in high posterior uncertainty.

While ASFRs represent the baseline data available in most settings, demographic analyses can sometimes draw upon richer sources of information. We therefore test how the parameter estimation process can be strengthened by incorporating such additional information. This can be done in two primary ways: either by augmenting the data with more detailed summary statistics or by encoding external knowledge into the model through more informative priors. Given that our model explicitly simulates the planned and unplanned nature of births, information on birth intentionality is a particularly relevant source of additional data. However, since this type of data is costly to collect and often unavailable, it is valuable to assess whether similar gains in estimation accuracy can be achieved by instead using informative priors on key behavioral parameters.

We therefore evaluate parameter recovery under three increasingly informative inference scenarios:

\textbf{Scenario 1: ASFRs with Weak Priors}: This baseline scenario uses only simulated age-specific fertility rates as summary statistics. All model parameters are assigned weakly informative priors.

\textbf{Scenario 2: ASFRs with Informative Priors.} This scenario tests the impact of incorporating plausible external information. The summary statistics are still limited to ASFRs, but we use narrower, more informative priors for the mean desired family size ($\mu_d$), the mean gap to intentional reproduction ($\delta_r$), and the mean desired birth spacing ($\mu_b$). All other parameters retain their weakly informative priors.

\textbf{Scenario 3: ASFRs and ASUFRs with Weak Priors.} This scenario tests the impact of adding more detailed data. We revert to the weakly informative priors from Scenario 1, but we augment the summary statistics to include both simulated ASFRs and age-specific unplanned fertility rates.

The informative priors used in Scenario 2 simulate a context where a researcher might leverage external information to improve estimates. For the mean desired family size ($\mu_d$), we construct the prior directly from the empirical survey distribution for the U.S. case, mimicking a situation where such summary data is readily available. For the remaining two parameters, we simulate the process of knowledge transfer from a data-rich to a data-poor setting by using the posteriors from our most data-intensive setup (Scenario 3) as the informative priors for Scenario 2. This mimics a situation where results from a detailed prior study on a similar cohort/country are used to inform a new analysis where only ASFRs are available.

\subsection{Prior Specification}

For our baseline scenario (Scenario 1), we selected weakly informative priors for all eleven model parameters to allow the data to drive the inference. The specific parameterization of these priors was chosen to cover a broad yet plausible range of values based on existing demographic literature and the inherent constraints of each parameter.

For strictly positive quantities like means and standard deviations (e.g., $\mu_s, \sigma_s$), we used broad Gamma distributions. For example, the prior for the mean age at sexual initiation ($\mu_s$) contains 95\% of its probability mass concentrated between 14.4 and 28.9 years. This range comfortably contains existing estimates for the mean age at onset of sexual activity across diverse Western countries and cohorts, which typically falls between 16 and 19 years \citep{wellings2006sexual, MacQuarrie2017}. Likewise, the prior for mean desired family size ($\mu_d$) is centered at 4.5 children, with a 95\% probability interval of [1.9, 8.1], again comfortably encompassing the available estimates for the cohorts studied \citep{westoff2015contraceptive, bongaarts1992fertility, sobotka2014two}. For parameters where reliable external estimates are lacking, such as the gap to intentional reproduction ($\delta_r$) and mean desired birth spacing ($\mu_b$), we employed Uniform priors to define very wide yet plausible ranges (0 to 8 years for $\delta_r$ and 10 to 100 months for $\mu_b$). The contraceptive failure probability ($\kappa$) was assigned a Beta distribution. Its prior reflects a wide range of plausible contraceptive efficacy levels for our study cohorts, centered around a 20\% failure rate but with significant probability assigned to both lower and higher values. Finally, the priors for the Bernstein coefficients ($\beta_1$ and $\beta_2$) were selected to ensure the resulting age-fecundability curves would generously contain available estimates of conception probabilities at early, peak, and later reproductive ages \citep{Schwartz1982,Dunson2002,wesselink2017age, ACOG2014}.

\section{Results}

This section presents the results from the validation experiments designed to test our framework under the three information scenarios defined above. We first assess the model's ability to recover known parameters from simulated data using a cross-validation procedure (Section \ref{sec:cv_results}). We then apply the framework to empirical data from our four study countries, evaluating both the in-sample fit through posterior predictive checks and, most critically, the out-of-sample predictive performance against micro-level data not used for estimation.

\subsection{Cross-Validation}
\label{sec:cv_results}

Our validation procedure involves repeatedly generating datasets with known parameters and then attempting to recover these parameters using our inference procedure. Specifically, we repeat the following steps for $I$ iterations:

\begin{enumerate}
    \item \textbf{Generate Ground-Truth Data:} In each iteration $i$ (for $i=1, \dots, I$), we draw a distinct "true" parameter vector, denoted $\theta^{*(i)}$, from the prior distribution $p(\theta)$ defined for the given scenario. Using our individual-level simulation model, we then generate a corresponding vector of aggregate summary statistics, denoted $x_0^{(i)}$, which serves as the pseudo-observed data for this iteration. The pair $(\theta^{*(i)}, x_0^{(i)})$ constitutes a ground-truth dataset where the data-generating parameters are known.

    \item \textbf{Perform Full Posterior Inference:} For each pseudo-observed dataset $x_0^{(i)}$, we independently run our complete Sequential Neural Posterior Estimation procedure. This involves training a neural density estimator over $R$ rounds with $N_{sim}$ new simulations drawn from the evolving proposal distribution in each round, to obtain an approximation of the posterior distribution $q_\phi(\theta \mid x_0^{(i)})$.

\item \textbf{Evaluate Parameter Recovery:} From each resulting posterior approximation $q_\phi(\theta \mid x_0^{(i)})$, we use the posterior mean as the point estimate for the parameters, denoted $\hat{\theta}^{(i)}$. We then assess the accuracy of recovery by comparing this estimate $\hat{\theta}^{(i)}$ with the known true generating parameters $\theta^{*(i)}$.
\end{enumerate}

By performing these $I$ independent inference runs, we obtain a distribution of estimation errors (e.g., $\hat{\theta}^{(i)} - \theta^{*(i)}$) for each model parameter under each of the three informational scenarios. We then summarize these error distributions using the normalized root-mean-squared error (RMSE). Lower RMSE values indicate better parameter recovery and allow for a direct comparison of how effectively each informational setup constrains the model parameters.

\begin{figure}[H]
  \centering
  \scriptsize
  \captionsetup{aboveskip=2pt, belowskip=2pt}

  \setlength{\tabcolsep}{0pt}
  \renewcommand{\arraystretch}{0.6}

  \newcommand{\rscat}[2]{%
    \raisebox{-0.5\height}{%
      \includegraphics[height=2.8cm,keepaspectratio]{plots/scatter_#1_#2.png}}}
  \newcommand{\rrmse}[1]{%
    \raisebox{-0.5\height}{%
      \includegraphics[height=2.8cm,keepaspectratio]{plots/normalized_rmse_#1.png}}}

  \scalebox{0.92}{%
    \begin{tabular}{@{}lcccc@{}}
        & Scenario 1 & Scenario 2 & Scenario 3 & Normalized RMSE\\[-0.5ex]

        $\mu_s$    & \rscat{01}{mu_s}    & \rscat{02}{mu_s}    & \rscat{03}{mu_s}    & \rrmse{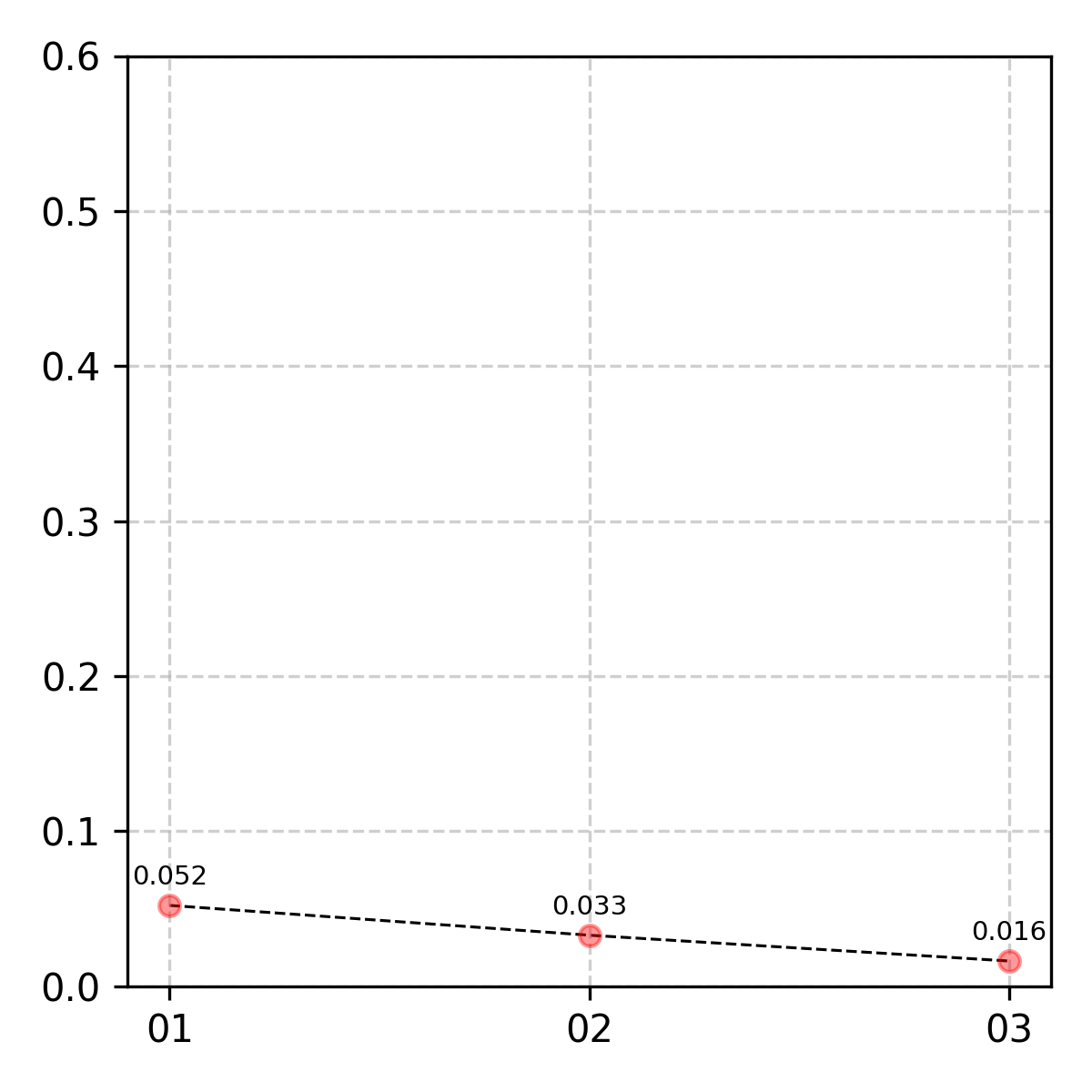}\\[-0.8ex]
        $\sigma_s$ & \rscat{01}{sigma_s} & \rscat{02}{sigma_s} & \rscat{03}{sigma_s} & \rrmse{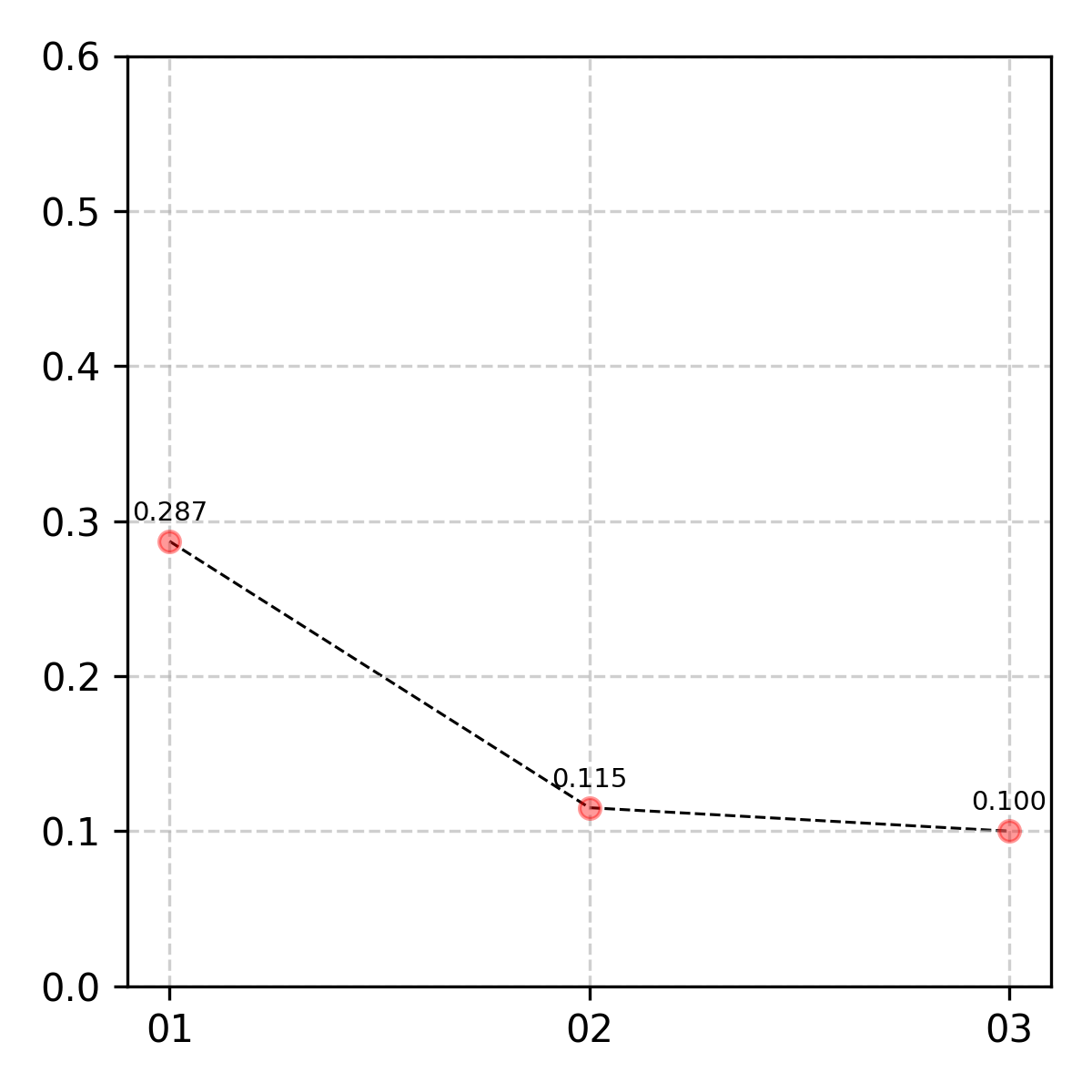}\\[-0.8ex]
        $\delta_r$ & \rscat{01}{delta_r} & \rscat{02}{delta_r} & \rscat{03}{delta_r} & \rrmse{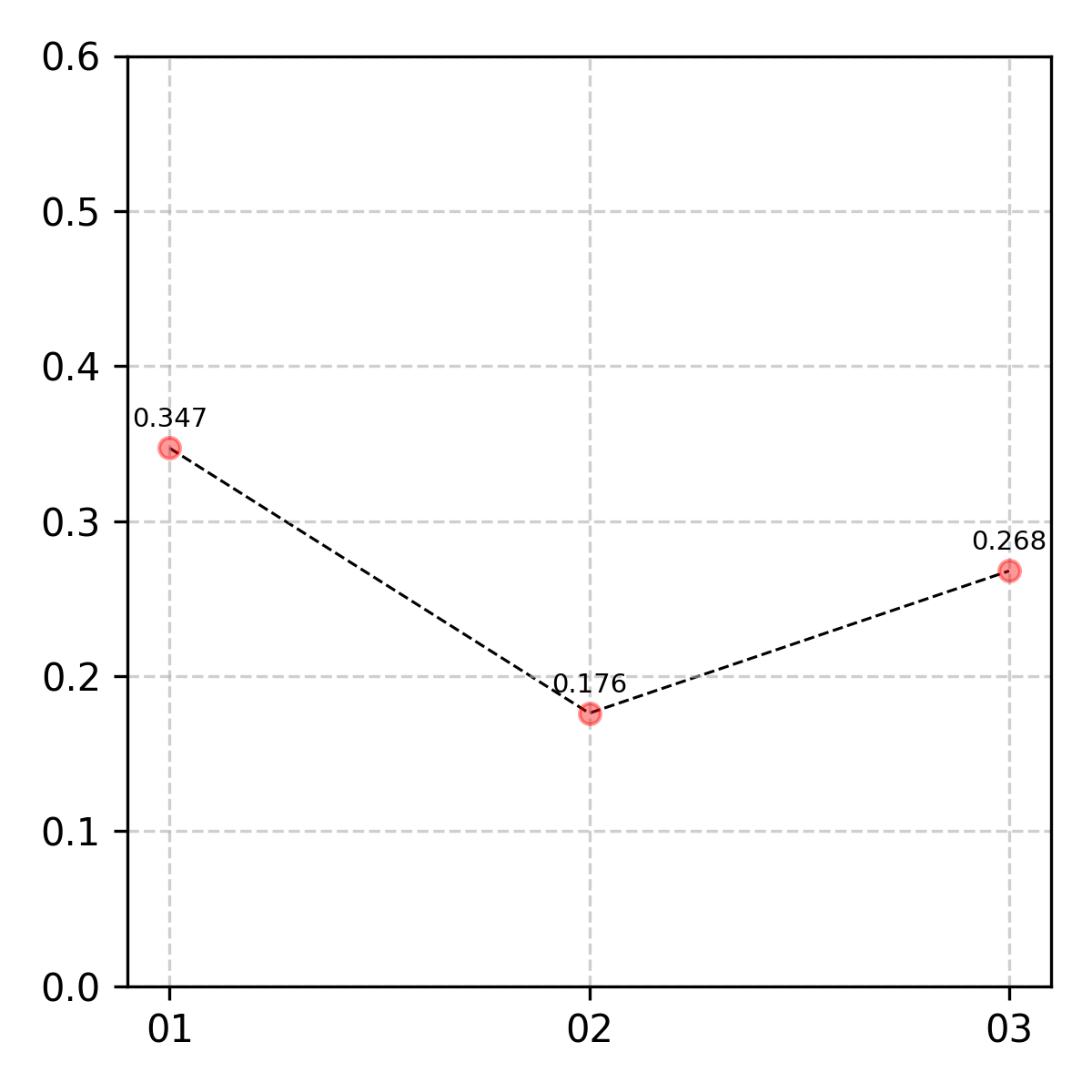}\\[-0.8ex]
        $\mu_d$    & \rscat{01}{mu_d}    & \rscat{02}{mu_d}    & \rscat{03}{mu_d}    & \rrmse{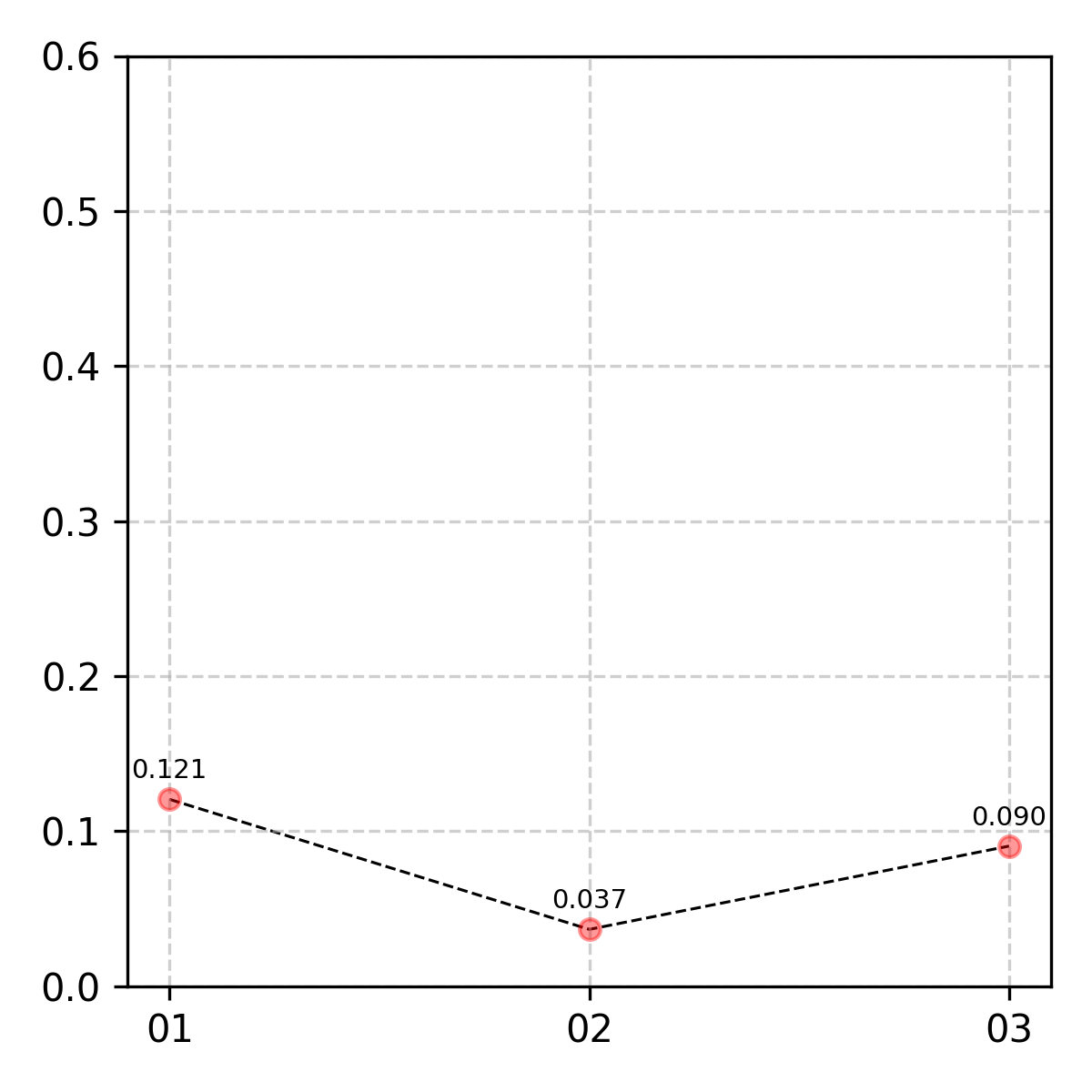}\\[-0.8ex]
        $\kappa$   & \rscat{01}{kappa}   & \rscat{02}{kappa}   & \rscat{03}{kappa}   & \rrmse{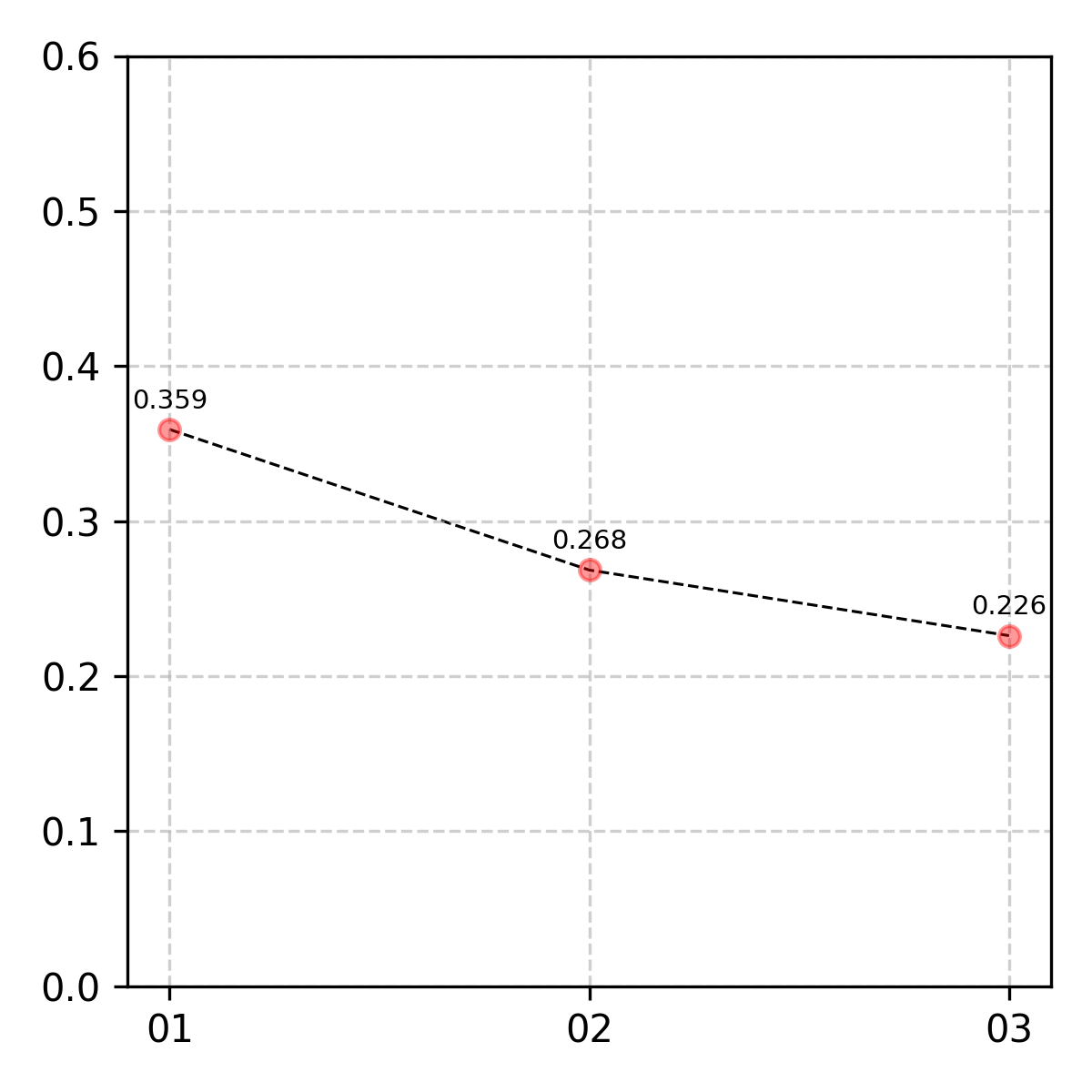}\\[-0.8ex]
        $\mu_b$    & \rscat{01}{mu_b}    & \rscat{02}{mu_b}    & \rscat{03}{mu_b}    & \rrmse{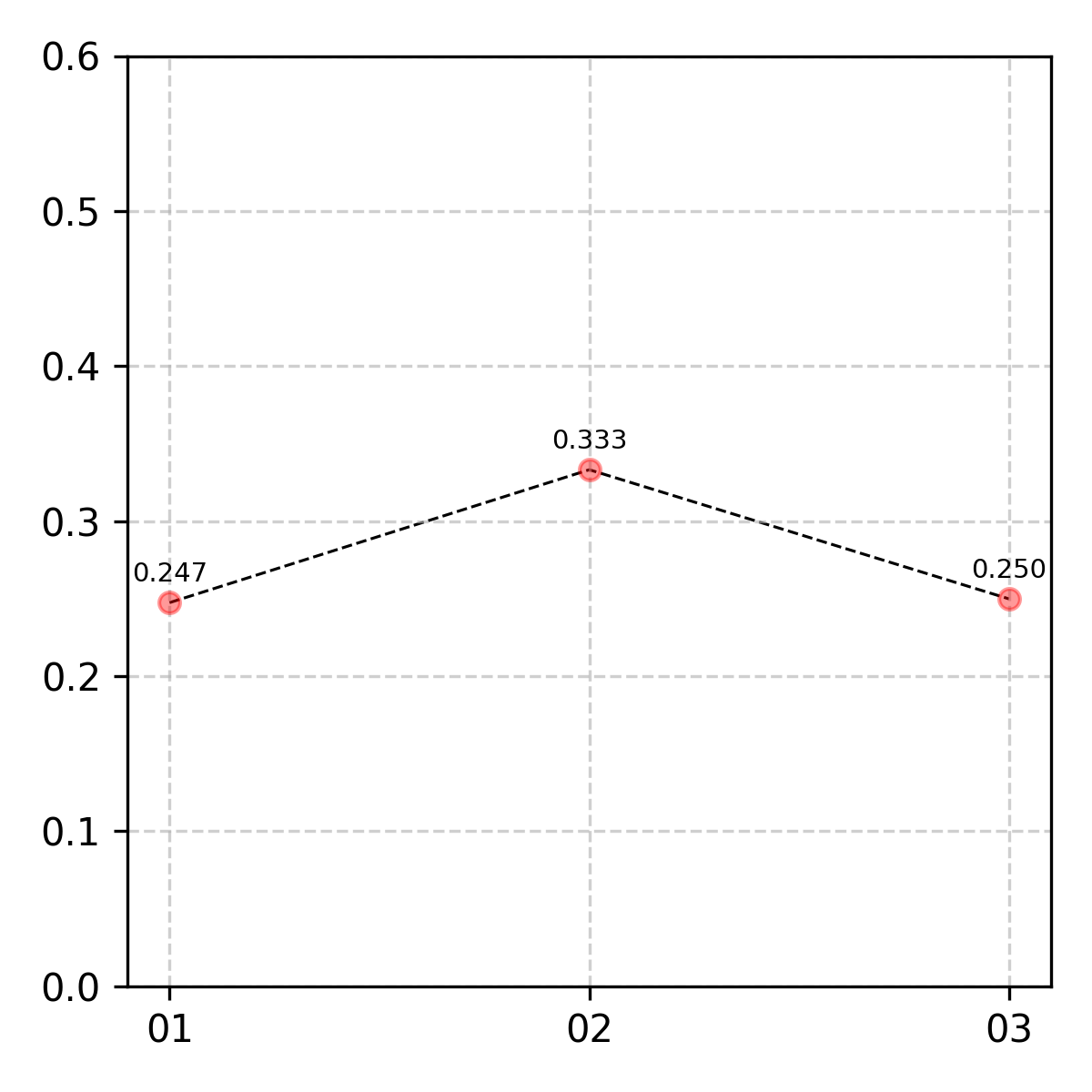}\\[-0.8ex]
        $\beta_1$  & \rscat{01}{beta1}   & \rscat{02}{beta1}   & \rscat{03}{beta1}   & \rrmse{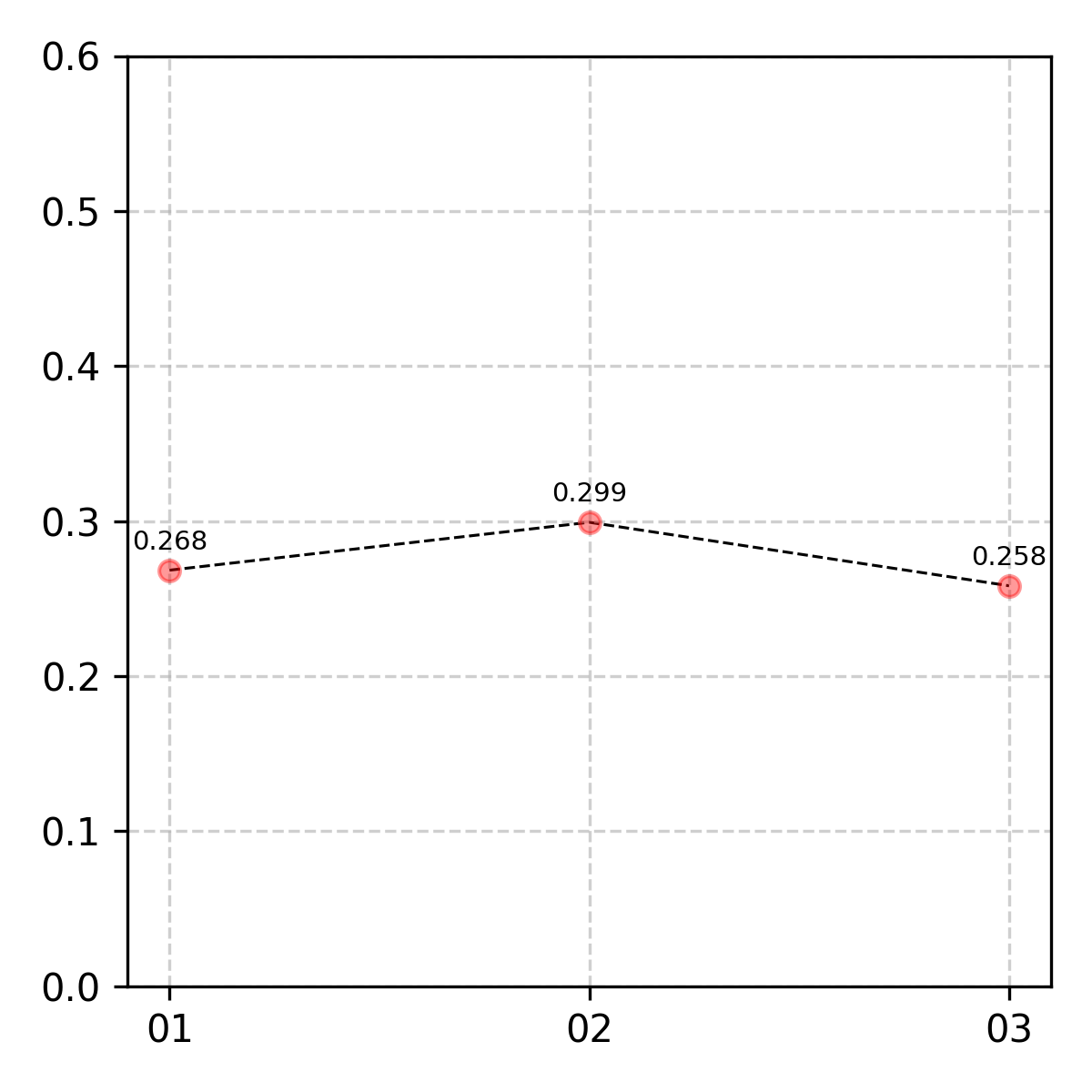}\\[-0.8ex]
        $\beta_2$  & \rscat{01}{beta2}   & \rscat{02}{beta2}   & \rscat{03}{beta2}   & \rrmse{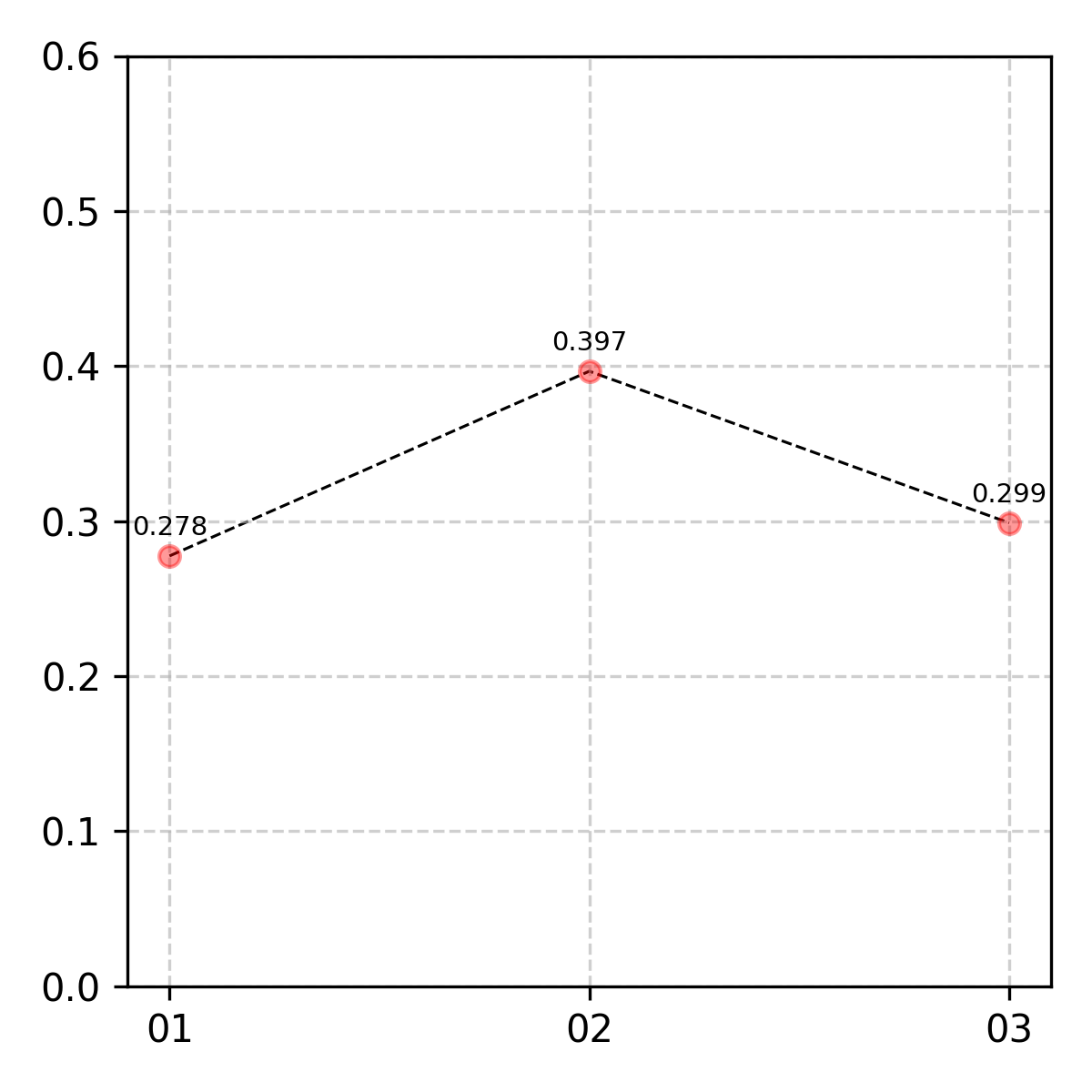}
    \end{tabular}%
  }

\caption{Cross‐validation over 25 folds. Columns 1–3 display scatterplots of true parameter values (horizontal axis) versus posterior mean estimates (vertical axis) for Scenarios 1, 2, and 3, with the dashed $45^\circ$ reference line indicating perfect recovery. Column 4 reports the scenario‐specific normalized root‐mean‐squared error.}
  \label{fig:cv_parameter_recovery}
\end{figure}

Figure~\ref{fig:cv_parameter_recovery} presents the results of our cross-validation analysis, performed over 25 folds for all eleven model parameters. The first three columns display scatterplots comparing the known "true" parameter values (horizontal axis) against the posterior mean estimates recovered by our model (vertical axis) for each of the three scenarios. Perfect parameter recovery would place all points directly on the dashed $45^\circ$ reference line. The final column summarizes the performance for each parameter by plotting the normalized root-mean-squared error (RMSE).

The results demonstrate that parameter recovery is already reasonably robust in the baseline scenario. In the scatterplots for this scenario, the estimated parameter values consistently cluster around the true values on the 45-degree line, with some parameters exhibiting tighter clustering than others. This indicates that the model can successfully disambiguate the majority of its behavioral components from the aggregate fertility schedule alone. This is a significant finding, as it was not obvious a priori that a single aggregate series could provide sufficient information to disentangle the rich set of mechanisms in our model. As expected, parameter recovery improves systematically as more information is incorporated, with the RMSE declining for nearly every parameter in Scenarios 2 and 3. 

\subsection{Model Validation}
\label{sec:validation}

The cross-validation experiments (Section~\ref{sec:cv_results}) confirmed the capacity of our SBI framework to recover key micro-level parameters from aggregate ASFRs. To further scrutinize the model's performance, we conduct two additional forms of validation. First, posterior predictive checks, to assesses the model's ability to replicate the aggregate data used for inference. The second validation stage examines the model's capacity to predict out-of-sample individual-level characteristics of the observed cohorts, such as distributions of desired family size, age at first sexual intercourse, and birth interval durations, which were not used during the parameter estimation process. 

We present the main validation results using the posteriors obtained under Scenario 1 (ASFRs with weak priors). We focus on this baseline scenario as it provides the most stringent test of our model's capacity to infer individual-level mechanisms solely from aggregate fertility rates. Results for the scenarios incorporating additional information (Scenarios 2 and 3) are presented in \hyperref[app:scenarios_full]{the Appendix}.

\subsubsection{Posterior Predictive Checks}
\label{sec:ppc}

To perform posterior predictive checks, we draw 5,000 samples from the joint posterior distribution for each of the four countries (United States, Colombia, Dominican Republic, and Peru). For each parameter sample, we simulate a full cohort of individual reproductive histories using our model and then recompute the ASFRs.

\begin{figure}[H]
  \centering
  \begin{subfigure}[b]{0.48\linewidth}
    \centering
    \includegraphics[width=\linewidth]{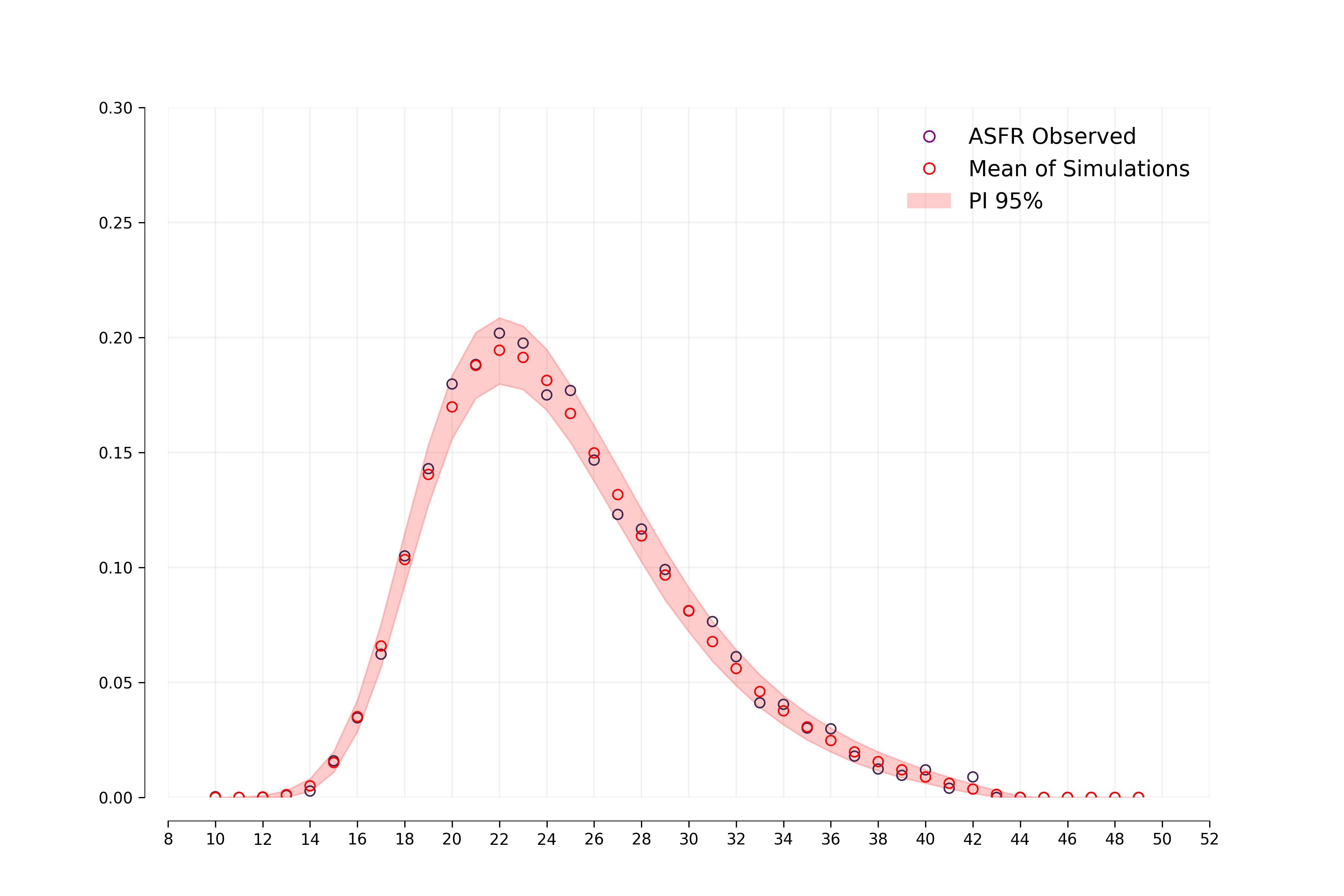}
    \caption{United States}
    \label{fig:ppc_US}
  \end{subfigure}
  \hfill
  \begin{subfigure}[b]{0.48\linewidth}
    \centering
    \includegraphics[width=\linewidth]{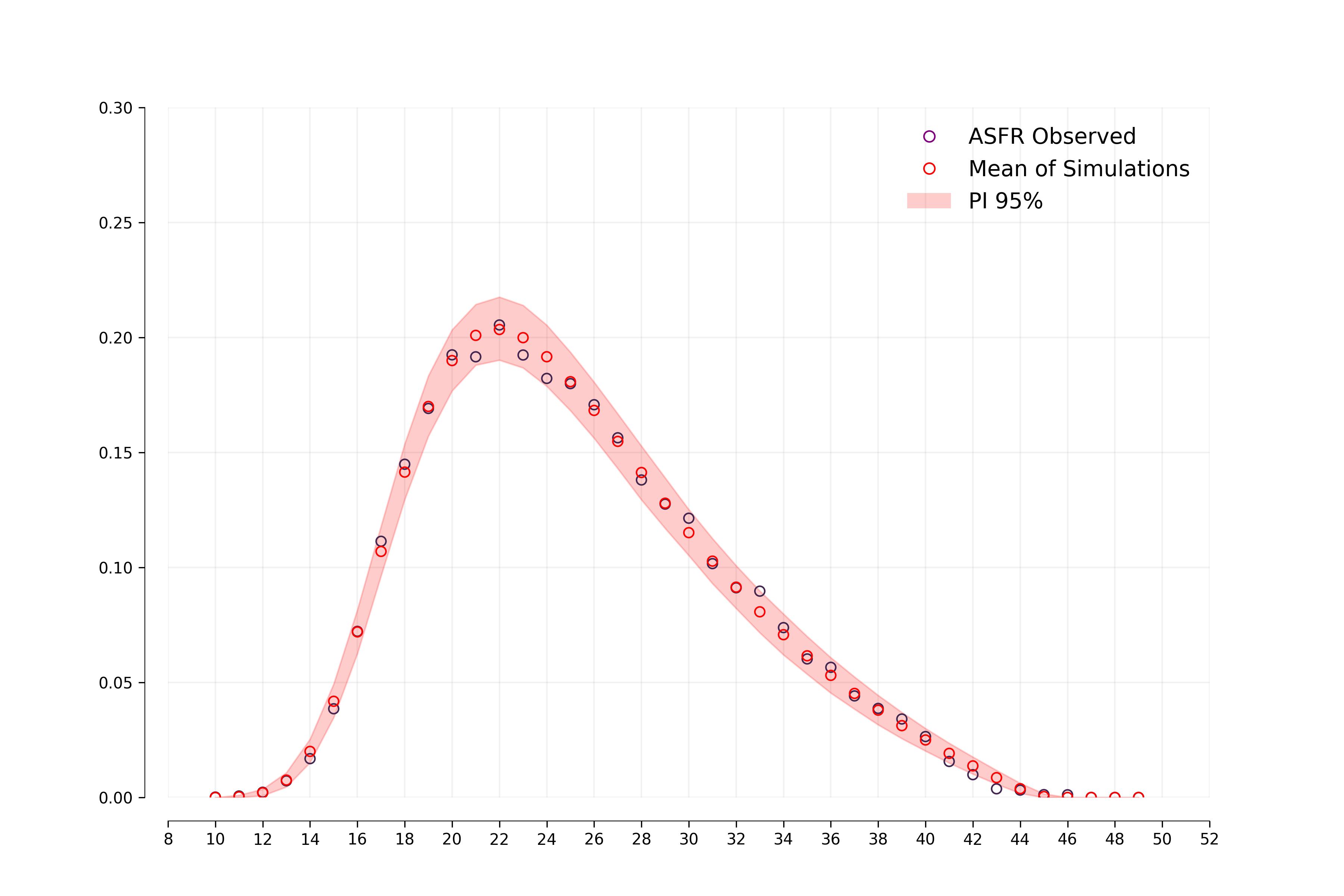}
    \caption{Colombia}
    \label{fig:ppc_CO}
  \end{subfigure}

  \vspace{0.5em}
  \begin{subfigure}[b]{0.48\linewidth}
    \centering
    \includegraphics[width=\linewidth]{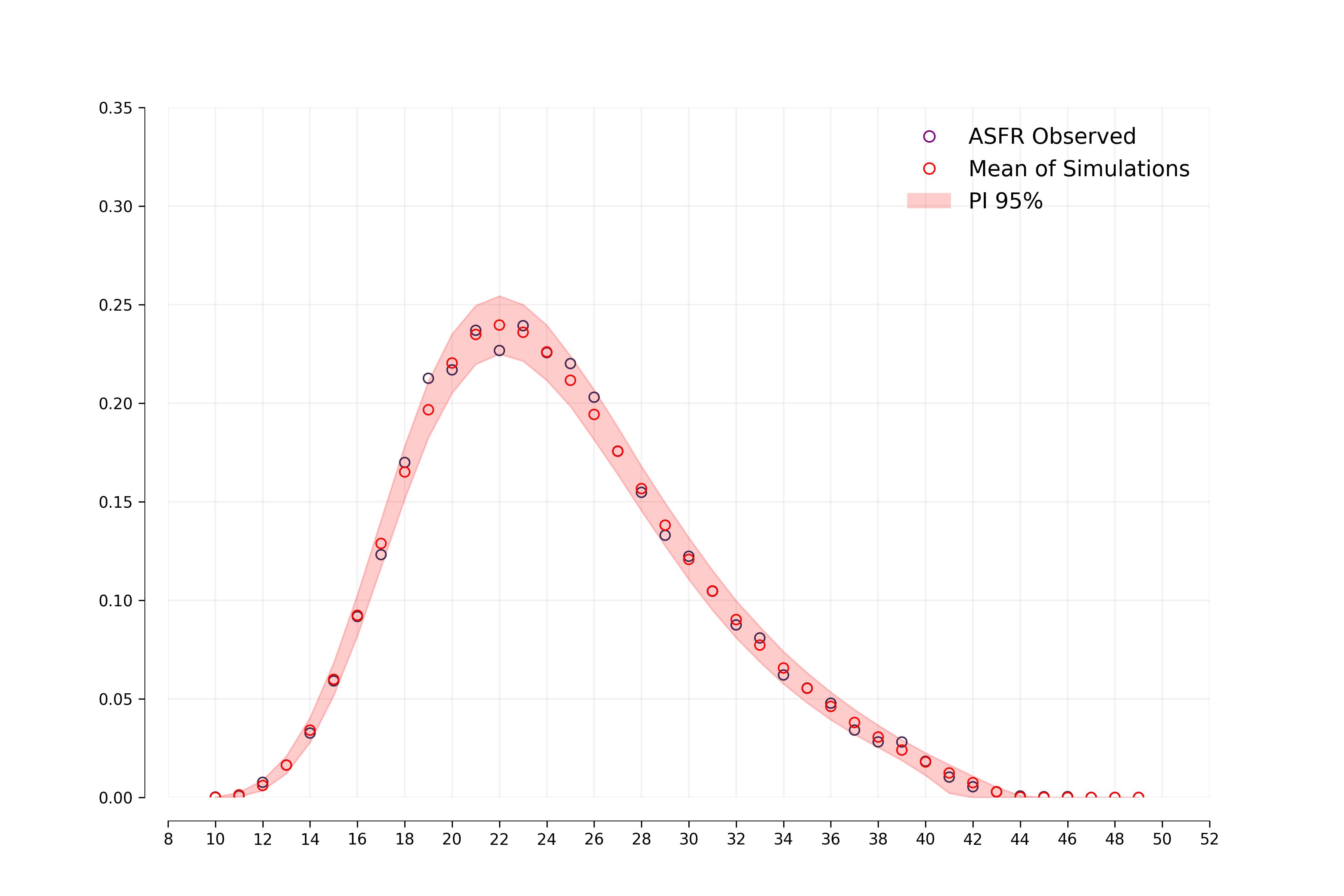}
    \caption{Dominican Republic}
    \label{fig:ppc_DR}
  \end{subfigure}
  \hfill
  \begin{subfigure}[b]{0.48\linewidth}
    \centering
    \includegraphics[width=\linewidth]{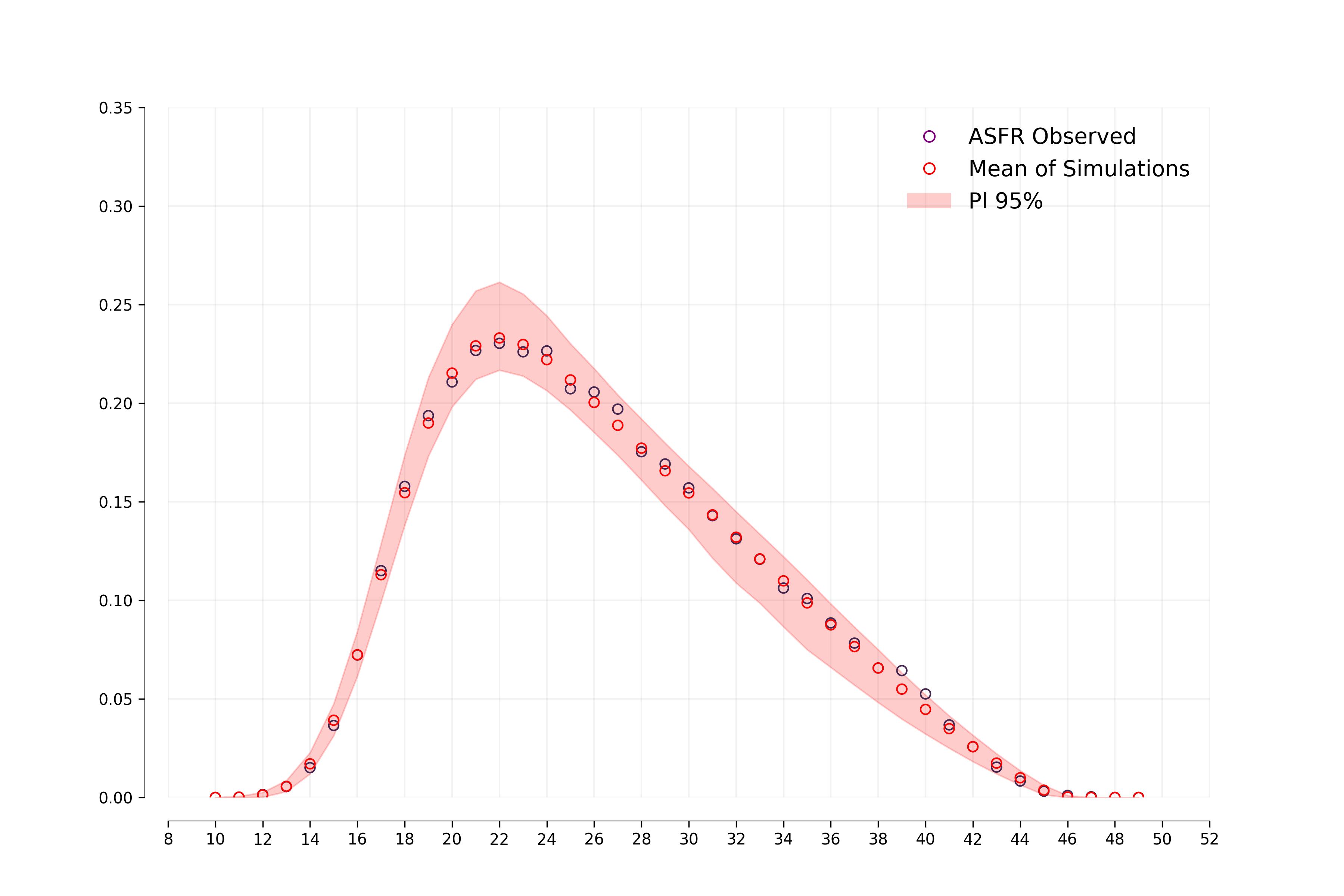}
    \caption{Peru}
    \label{fig:ppc_PE}
  \end{subfigure}

	\caption{\textbf{Posterior Predictive Checks for Age-Specific Fertility Rates.} Comparison of observed rates (purple open circles) with posterior predictions for four countries: (a) United States, (b) Colombia, (c) Dominican Republic, and (d) Peru. Red open circles indicate the mean of 5,000 posterior predictive simulations. Red shaded areas represent the 95\% posterior predictive intervals. Parameters were estimated under Scenario 1 (ASFRs with weakly informative priors).}
	\label{fig:ppc_plots}
\end{figure}

Figure~\ref{fig:ppc_plots} presents observed ASFRs against their corresponding posterior predictive distributions. Observed ASFRs are represented by purple open circles. Red open circles represent the mean of the 5,000 simulated ASFRs schedules, while the shaded red areas depict the 95\% posterior predictive intervals. 

Across all four diverse demographic contexts, the model demonstrates a strong capacity to replicate the observed data. The posterior predictive means closely track the observed ASRFs, capturing key features such as the location and height of peak fertility, and the age pattern of decline. Notably, across all countries, almost all observed points fall well within the 95\% posterior predictive interval, indicating that the model, conditioned on the inferred parameters, generates aggregate outcomes highly consistent with the observed data.

These results confirm that our framework, even with minimal data inputs, can consistently and accurately reproduce the population-level fertility schedules it was intended to explain. 

\subsection{Estimated Marginal Posteriors}

Figure~\ref{post_nat} displays the marginal posterior distributions for all eleven model parameters, estimated under Scenario 1 for each of the four countries. The results clearly show that the aggregate ASFRs data provide substantial information for constraining most parameters. In nearly all cases, the posterior distributions (in red) are considerably sharper and more concentrated than their corresponding wide priors (in grey), indicating a significant reduction in uncertainty after conditioning on the observed data.

The degree to which parameters are constrained varies logically across the different components of the model. Parameters related to the timing of sexual initiation ($\mu_s, \sigma_s$), the age-pattern of fecundability ($\beta_1, \beta_2$), desired family size ($\mu_d, \sigma_d$), and contraceptive failure ($\kappa$) are all well-constrained, exhibiting sharp, unimodal posterior distributions. The posteriors also reveal plausible cross-national differences. For example, the mean desired family size ($\mu_d$) for the U.S. cohort is lower than in the other three countries, as expected. Similarly, the contraceptive failure probability ($\kappa$) is estimated to be lowest in the U.S., a finding that aligns with the socio-demographic context of these cohorts. In contrast, parameters governing the timing between life-course events, like the gap to intentional reproduction ($\delta_r$) and the spacing between births ($\mu_b, \sigma_b$), show higher posterior uncertainty. Their posteriors are wider and less distinguished from their priors, suggesting that while ASFRs alone are sufficient to identify the core components of the model, pinpointing the precise timing of \emph{intended} fertility is more challenging without additional data on birth intentionality. However, it is important to note that this remaining uncertainty in specific parameters does not significantly degrade the model's overall predictive performance, as evidenced by the successful posterior predictive checks (Figure~\ref{fig:ppc_plots}). Finally, the estimation for these timing-related parameters improves substantially when direct information on birth intentionality is included, as shown in the results for Scenario 3 (see Figure~\ref{fig:appendix_posterior_comparison} in \hyperref[app:scenarios_full]{the Appendix}).

\floatsetup[figure]{capposition=bottom}
\setlength\tabcolsep{4pt}
\renewcommand{\arraystretch}{1.05}
\adjustboxset{width=\linewidth,valign=c}
\newcommand{\plot}[2]{%
  \includegraphics[width=0.8\linewidth,keepaspectratio]{plots/#1_#2.pdf}%
}

\begin{figure}[H] 
  \centering
  \begin{tabularx}{\linewidth}{@{}l*{4}{X}@{}}
    & \multicolumn{1}{c}{\textbf{US}}
    & \multicolumn{1}{c}{\textbf{CO}}
    & \multicolumn{1}{c}{\textbf{DR}}
    & \multicolumn{1}{c}{\textbf{PE}} \\[2pt]

    \textbf{$\mu_s$} & \plot{mu_s}{US} & \plot{mu_s}{CO} & \plot{mu_s}{DR} & \plot{mu_s}{PE} \\
    \textbf{$\sigma_s$} & \plot{sigma_s}{US} & \plot{sigma_s}{CO} & \plot{sigma_s}{DR} & \plot{sigma_s}{PE} \\
    \textbf{$\delta_r$} & \plot{delta_r}{US} & \plot{delta_r}{CO} & \plot{delta_r}{DR} & \plot{delta_r}{PE} \\
    \textbf{$\sigma_r$} & \plot{sigma_r}{US} & \plot{sigma_r}{CO} & \plot{sigma_r}{DR} & \plot{sigma_r}{PE} \\
    \textbf{$\mu_d$} & \plot{mu_d}{US} & \plot{mu_d}{CO} & \plot{mu_d}{DR} & \plot{mu_d}{PE} \\
    \textbf{$\sigma_d$} & \plot{sigma_d}{US} & \plot{sigma_d}{CO} & \plot{sigma_d}{DR} & \plot{sigma_d}{PE} \\
  \end{tabularx}
  \refstepcounter{figure}
\end{figure}

\begin{figure}[H]
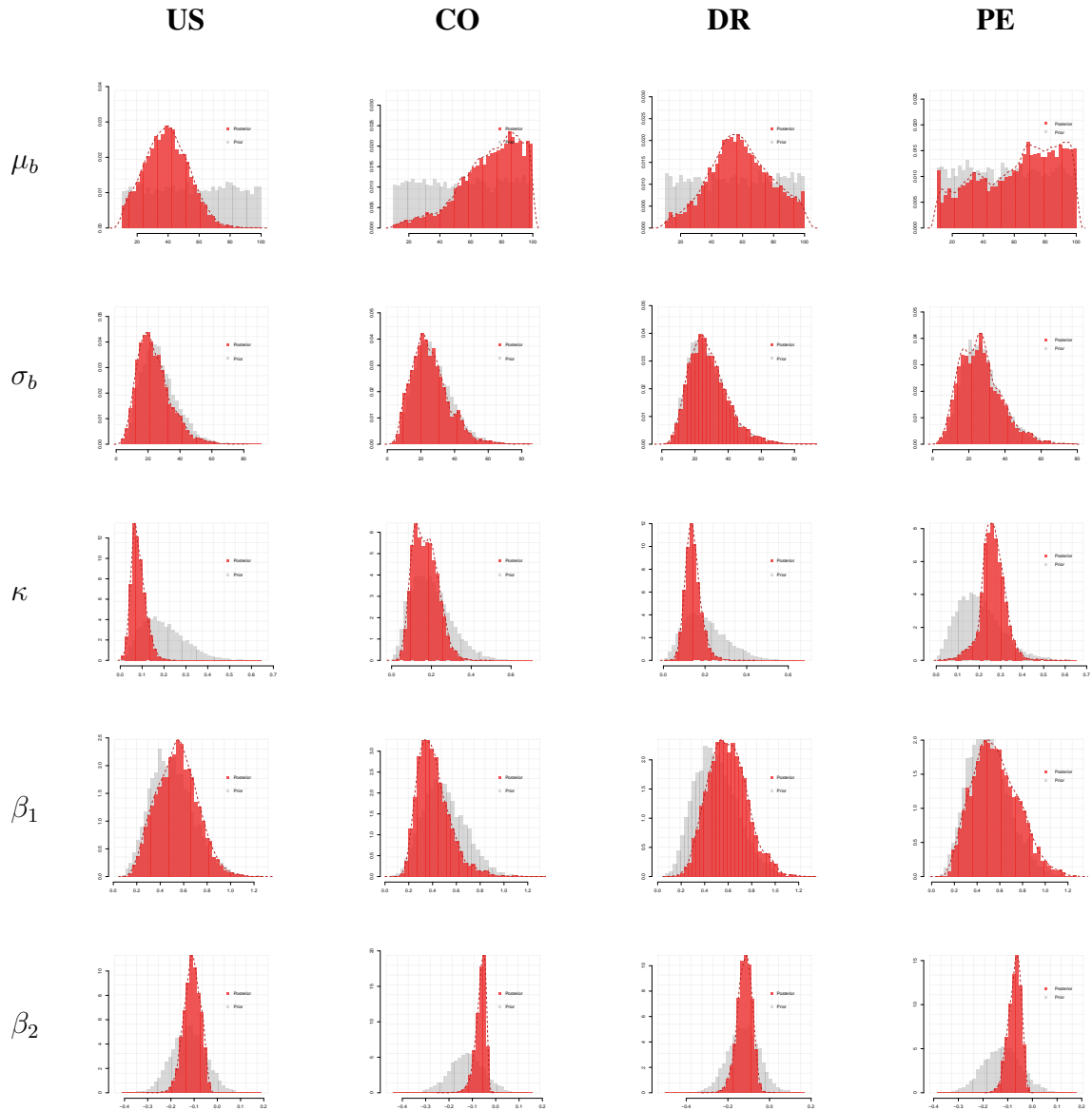
 
  \ContinuedFloat 
  \centering
  \begin{tabularx}{\linewidth}{@{}l*{4}{X}@{}}
    & \multicolumn{1}{c}{\textbf{US}}
    & \multicolumn{1}{c}{\textbf{CO}}
    & \multicolumn{1}{c}{\textbf{DR}}
    & \multicolumn{1}{c}{\textbf{PE}} \\[2pt]

    \textbf{$\mu_b$} & \plot{mu_b}{US} & \plot{mu_b}{CO} & \plot{mu_b}{DR} & \plot{mu_b}{PE} \\
    \textbf{$\sigma_b$} & \plot{sigma_b}{US} & \plot{sigma_b}{CO} & \plot{sigma_b}{DR} & \plot{sigma_b}{PE} \\
    \textbf{$\kappa$} & \plot{kappa}{US} & \plot{kappa}{CO} & \plot{kappa}{DR} & \plot{kappa}{PE} \\
    \textbf{$\beta_{1}$} & \plot{beta1}{US} & \plot{beta1}{CO} & \plot{beta1}{DR} & \plot{beta1}{PE} \\
    \textbf{$\beta_{2}$} & \plot{beta2}{US} & \plot{beta2}{CO} & \plot{beta2}{DR} & \plot{beta2}{PE} \\
  \end{tabularx}
  \caption{Marginal posterior distributions of the eleven key model parameters for the United States, Colombia, Dominican Republic, and Peru, inferred from empirical data under Scenario 1.}
  \label{post_nat}
\end{figure}

\subsubsection{Out‐of‐Sample Validation}
\label{sec:out}

\begin{figure}[H]
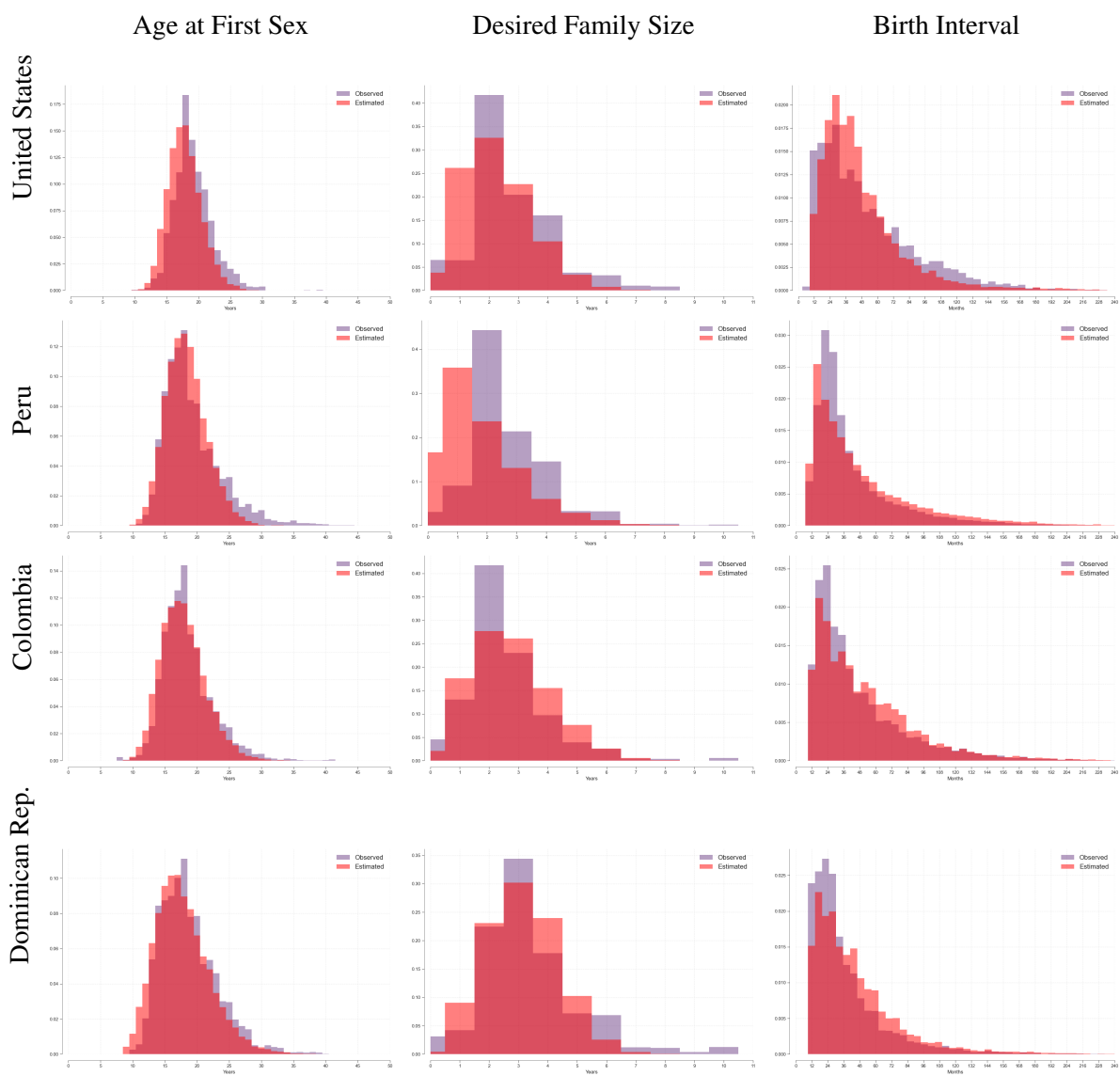

  \centering
  \small

  \newcommand{\valplot}[2]{%
    \includegraphics[width=.32\linewidth]{plots/validation_#2_#1_s1.png}}

  \setlength{\tabcolsep}{2pt} 
  \renewcommand{\arraystretch}{1.0}
  \begin{tabular}{@{}c ccc@{}}
    & \small Age at First Sex & \small Desired Family Size & \small Birth Interval\\[4pt]

    \rotatebox{90}{\small United States} &
      \valplot{US}{age_start_sex} &
      \valplot{US}{desired_family_size} &
      \valplot{US}{birth_intervals}\\[4pt]

    \rotatebox{90}{\small Peru} &
      \valplot{PE}{age_start_sex} &
      \valplot{PE}{desired_family_size} &
      \valplot{PE}{birth_intervals}\\[4pt]

    \rotatebox{90}{\small Colombia} &
      \valplot{CO}{age_start_sex} &
      \valplot{CO}{desired_family_size} &
      \valplot{CO}{birth_intervals}\\[4pt]

    \rotatebox{90}{\small Dominican Rep.} &
      \valplot{DR}{age_start_sex} &
      \valplot{DR}{desired_family_size} &
      \valplot{DR}{birth_intervals}
  \end{tabular}

  \caption{\textbf{Validation on micro-level outcomes.} Purple bars denote the observed survey distributions, while red bars show the model-implied counterparts generated from posterior draws. Columns display key behavioral dimensions: age at first sex, desired family size, and birth interval, whereas rows correspond to each study cohort.}
  \label{fig:validation}
\end{figure}

To assess the model's ability to generate realistic individual-level reproductive behaviors not directly used in parameter estimation, we use parameter values representative of the posterior distribution (specifically, the posterior mean obtained under Scenario 1, as described in Section~\ref{sec:validation}) to drive new simulations. These simulations yield full synthetic reproductive histories. We then compare the distributions of key micro-level outcomes from these synthetic life courses, namely, age at first sexual activity, desired family size, and inter-birth intervals, with their empirical counterparts derived from the survey data (Figure~\ref{fig:validation}). This side-by-side comparison provides a direct appraisal of the model's realism in capturing individual-level reproductive patterns.

\paragraph{Overall fit.}

Across the four countries and three micro-level indicators, the distributions generated from the model's synthetic life courses generally provide a strong approximation of the empirical distributions observed in the survey data. This indicates that the parameters inferred solely from aggregate ASFRs enable the reconstruction of distinct micro-behavioral dimensions without any specific tuning towards these individual-level targets. Quantitatively, with the notable exception of desired family size in Peru, the Jensen-Shannon divergence (base 2) between simulated and observed distributions never exceeds 0.068 bits.

\paragraph{Age at first sex.}
The simulated distributions reproduce the overall shape and cross-country ordering of sexual debut: the mean age is highest in the United States and Peru, slightly lower in Colombia, and lowest in the Dominican Republic. The model naturally smooths over sharp heaping in the survey data (e.g., at age 18), and for the United States, the predicted curve is displaced slightly to the left of the observed one. Even with these minor mismatches, the alignment is striking, given the relative simplicity of the model and of the data used for its estimation.

\paragraph{Desired family size.}
Overall, the simulated preferences closely match the shape of the observed distributions, reproducing the main peak around two to three children and the gradual taper into larger family ideals. A clear deviation appears at exactly two children, where the empirical data show a sharp bump (a clear outcome of strong social norms) that the model's simple two-parameter form inevitably smooths over. A more significant discrepancy is observed for Peru, where the model's predicted distribution is shifted to the left of the empirical one (JS divergence: 0.144 bits). This specific mismatch is consistent with the higher posterior uncertainty observed for the Peruvian cohort and may reflect underlying issues like lower data quality or greater-than-modeled heterogeneity in reproductive behavior in this cohort. It is important to note, however, that the alignment of these distributions improves substantially when additional information is incorporated (see Figure~\ref{fig:appendix_validation_dfs} in the Appendix).

\paragraph{Birth intervals.}
The synthetic data recovers the principal spacing pattern: a broad peak around two to three years followed by a gradual decline towards longer gaps. The model undershoots the cluster of very short gaps and shows a slightly heavier tail beyond five years. Apart from these edges of the distribution, the match is close across the main mass of birth intervals.

\section{Discussion}
\label{sec:discussion}

This study has introduced a likelihood-free Bayesian framework capable of inferring complex individual-level reproductive behaviors solely from aggregate population data. By integrating a demographically interpretable individual-level simulation model with Sequential Neural Posterior Estimation, we demonstrated that the core parameters governing the reproductive process can be recovered from age-specific fertility rates alone. While incorporating additional information (either through the model priors or more detailed data) serves to further reduce parameter uncertainty, our findings show that the information contained in ASFRs alone provide a robust baseline estimation. The model's ability to accurately reproduce observed fertility schedules, as confirmed by our posterior predictive checks, offers strong support for this conclusion. However, the most compelling validation of our approach lies in the model's ability, using these inferred parameters, to accurately predict out-of-sample micro-level distributions for behaviors such as age at first sexual intercourse, desired family size, and birth intervals, none of which informed the original estimation. This establishes a new, statistically rigorous bridge between macro-level fertility observations and their underlying micro-behavioral drivers.

Methodologically, this work serves as an empirical demonstration of SNPE's utility for parameter inference in interpretable, yet complex, social science simulation models where direct likelihood calculation is prohibitive. The ability to recover these parameters from a single, widely available data series like ASFRs offers transferable insights for simulation-based inference across other disciplines facing similar inverse problems. Substantively, our framework provides a novel toolkit for demographers, enabling the estimation of behaviorally meaningful parameters (e.g., mean desired family size, age at intentional reproduction, contraceptive failure rates) even in data-scarce contexts, thereby offering a mechanistic complement to traditional micro-data analyses.

This capability to infer interpretable behavioral parameters significantly enhances population forecasting, moving beyond traditional methods that often rely on extrapolating aggregate trends or employ macro-level parameters with limited behavioral grounding. Our approach enables the generation of 'behaviorally explicit population forecasts' where future scenarios are constructed from hypothesized changes in these underlying, micro-level behavioral parameters. Crucially, by first distilling complex fertility dynamics into these more causally proximate behavioral components, our framework also facilitates a clearer understanding of how external covariates, such as education or economic conditions, shape reproductive outcomes. Machine learning models, for instance, may more effectively identify systematic relationships when predicting these interpretable behavioral parameters rather than attempting to directly predict highly aggregated and confounded ASFRs. This improved ability to model behavior mechanisms provides more robust and theoretically sound inputs for scenario-based population projections. Furthermore, by reducing the data requirements for microsimulation, our framework makes the construction of 'demographic digital twins' a more accessible and feasible tool for policy exploration.

Despite these advances, the present framework has limitations that define important avenues for future development. First, the individual-level model, while behaviorally detailed, necessarily simplifies the complex dynamics of human reproduction. Second, the data generating process encoded by our model is suited to contemporary populations where reproduction is substantially decoupled from sexuality and where such articulable birth control decision-making is prevalent. This may limit its direct applicability to contexts with more natural fertility regimes or where reproductive agency and intentions are structured differently. Third, the framework as presented assumes relatively homogeneous cohorts. Applying the model to highly heterogeneous populations where reproductive behavior differs systematically by socioeconomic characteristics would likely require disaggregated input data. A final consideration is that while ASFRs provide a robust foundation for estimation, applications that demand higher precision or focus on specific mechanisms may require supplementary data that are not always easily available.

These limitations point toward several exciting directions for future work. First, strategies can be developed to enhance the precision and utility of the framework, particularly for analyses relying solely on ASFRs. One important avenue is the incorporation of stronger, empirically-grounded priors on the key behavioral parameters, allowing researchers to leverage existing knowledge from demographically similar populations. Another promising direction is the development of methods to generate plausible supplementary data; for instance, one could use the learned associations between covariates (e.g., education, age) and birth intentionality from available survey to produce plausible estimates in data-scarce contexts. A second major direction involves adapting the framework’s core structure to broaden its applicability and complexity. A key extension would be to explicitly model population heterogeneity. In the many contexts where data such as education-specific ASFRs are available, the framework could be adapted to infer distinct behavioral parameters for different subgroups. For populations closer to natural fertility, the model could be simplified to focus on a smaller set of parameters, while for contexts where Western notions of "unplanned" births are less salient, research could explore identifying and incorporating alternative summary statistics that better reflect local modes of fertility regulation.

\pagebreak

\phantomsection
\section*{Appendix}
\label{app:scenarios_full}

\begin{figure}[H]
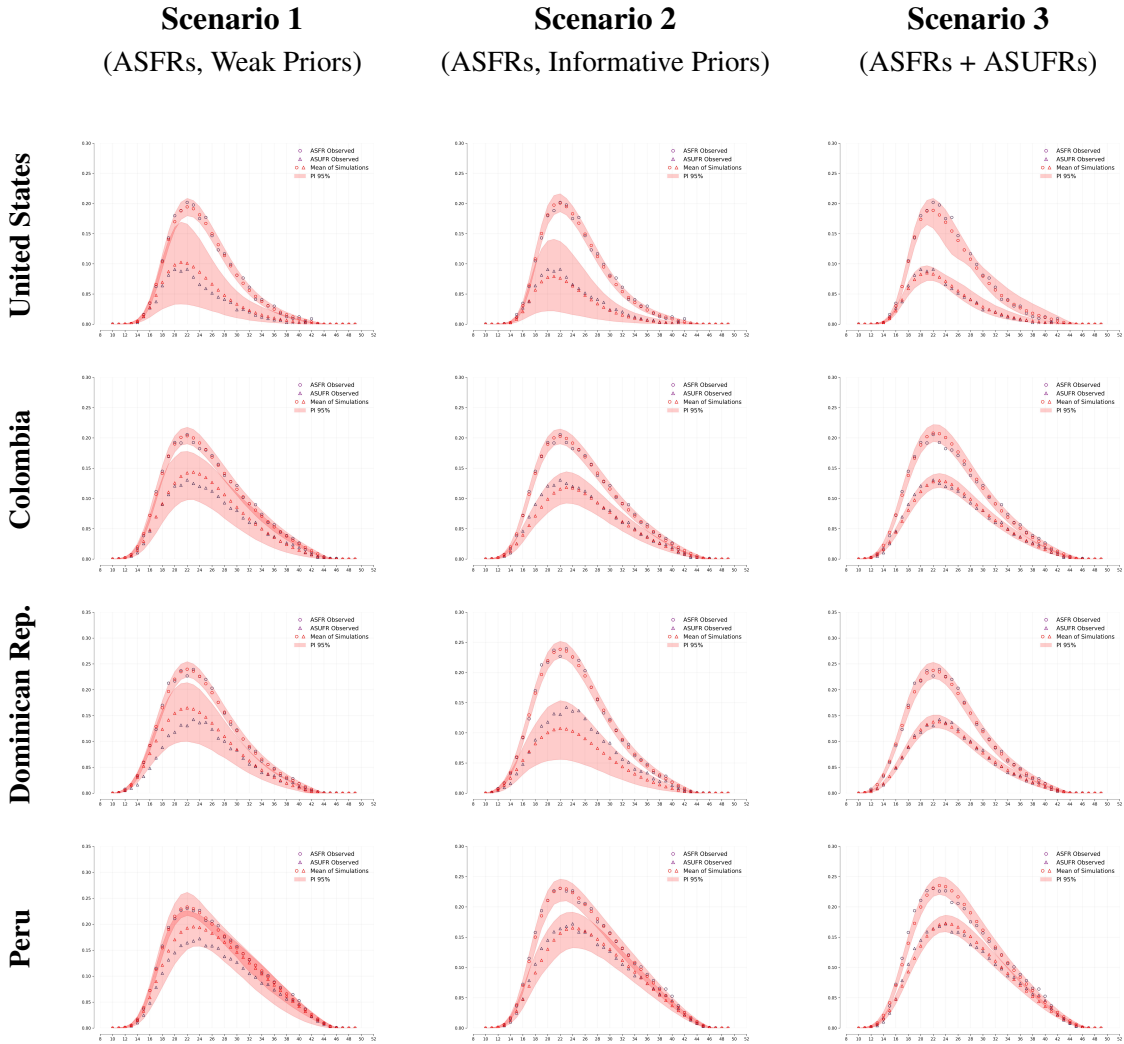

    \centering
    
    \newcommand{\ppcplot}[2]{%
        \includegraphics[width=0.3\linewidth]{plots/ppc_#1_s#2.jpg}%
    }

    \setlength{\tabcolsep}{4pt} 
    \begin{tabular}{lccc}
        & \textbf{Scenario 1} & \textbf{Scenario 2} & \textbf{Scenario 3} \\
        & \small(ASFRs, Weak Priors) & \small(ASFRs, Informative Priors) & \small(ASFRs + ASUFRs) \\[1em] 
        
        \raisebox{-.5\height}{\rotatebox{90}{\small\textbf{United States}}} &
        \ppcplot{US}{1} & \ppcplot{US}{2} & \ppcplot{US}{3} \\[1em]
        
        \raisebox{-.5\height}{\rotatebox{90}{\small\textbf{Colombia}}} &
        \ppcplot{CO}{1} & \ppcplot{CO}{2} & \ppcplot{CO}{3} \\[1em]

        \raisebox{-.5\height}{\rotatebox{90}{\small\textbf{Dominican Rep.}}} &
        \ppcplot{DR}{1} & \ppcplot{DR}{2} & \ppcplot{DR}{3} \\[1em]

        \raisebox{-.5\height}{\rotatebox{90}{\small\textbf{Peru}}} &
        \ppcplot{PE}{1} & \ppcplot{PE}{2} & \ppcplot{PE}{3} \\
    \end{tabular}

    \caption{Full posterior predictive checks for both Age-Specific Fertility Rates (ASFRs) and Unplanned Fertility Rates (ASUFRs) across all three inference scenarios. Each panel compares the observed rates with the 95\% posterior predictive interval generated under a given scenario (columns) for each country (rows).}
    \label{fig:appendix_ppc_full}
\end{figure}

Figure~\ref{fig:appendix_ppc_full} presents the full posterior predictive checks for all scenarios, now including the fit to the Age-Specific Unplanned Fertility Rate (ASUFR) alongside the ASFRs. While the model provides a strong fit to the observed ASFRs across all scenarios, the varying fit to the ASUFRs allows for a clearer visualization of the performance gains from incorporating different sources of information. Notably, even in the baseline scenario which uses no data on birth intentionality, the posterior predictive mean tracks the ASUFRs, although the associated predictive interval is considerably wide. This predictive uncertainty is progressively reduced across the subsequent scenarios. The use of informative priors in Scenario 2 reduces the distance between the observed data and the predictive posterior means and leads to a sharper and more precise predictive distribution. The fit becomes most precise in Scenario 3, where the inclusion of ASUFR data allows the model to replicate the observed pattern of unplanned fertility with high accuracy.

Figure~\ref{fig:appendix_posterior_comparison} compares the marginal posterior distributions across the three inference scenarios for each of the eleven model parameters and the four study countries (United States, Colombia, Dominican Republic, and Peru). In each panel, the posteriors for Scenario 1 (light purple), Scenario 2 (orange), and Scenario 3 (red) are overlaid, illustrating the progressive reduction in parameter uncertainty as additional information is incorporated. Overall, the posteriors become narrower and more peaked from Scenario 1 to Scenario 3, reflecting improved precision with richer inputs, though the extent of this sharpening varies by parameter and country.

Notable patterns emerge across scenarios. For parameters related to the timing of sexual initiation (e.g., $\mu_s$ and $\sigma_s$) and fecundability ($\beta_1$ and $\beta_2$), the posteriors are already relatively well-constrained in Scenario 1 but show further refinement in Scenarios 2 and 3, with minimal overlap between the distributions. In contrast, timing-related parameters like the gap to intentional reproduction ($\delta_r$) and desired birth spacing ($\mu_b$ and $\sigma_b$) exhibit higher initial uncertainty in Scenario 1, often with wider and flatter posteriors; this uncertainty decreases markedly in Scenarios 2 and 3, where the added information (priors or data) leads to substantial narrowing. The contraceptive failure rate ($\kappa$) and desired family size parameters ($\mu_d$ and $\sigma_d$) demonstrate consistent narrowing, particularly benefiting from the added information in Scenario 3, which reduces posterior variance by emphasizing unplanned fertility dynamics.

Differences in estimation behavior are evident across countries/datasets. For instance, the United States shows the most consistent posteriors across scenarios, with distributions narrowing progressively but remaining in similar locations; this stability is noteworthy given the higher data quality in the U.S. surveys, which may contribute to robust identifiability even under weaker priors. In contrast, Colombia and Peru exhibit more pronounced shifts and sharpening from Scenario 1 to the others, with posteriors in Scenario 1 often broader and sometimes multimodal, suggesting greater sensitivity to additional information in these contexts. The Dominican Republic falls in between, with moderate improvements in precision but less marked relocation of posterior mass. 

\newcommand{\postplot}[2]{%
    \includegraphics[width=\linewidth]{plots/posterior_comparison_#1_#2.png}%
}

\renewcommand\tabularxcolumn[1]{>{\centering\arraybackslash}p{#1}} 

\setlength{\LTcapwidth}{\textwidth} 

\renewcommand\LTcaptype{figure} 

\begin{longtable}{@{} l *{4}{ >{\centering\arraybackslash} p{0.23\textwidth} } @{} } 

\caption[Comparison of marginal posterior distributions across the three inference scenarios. Each panel shows the posterior for a given parameter (rows) and country (columns). The distributions for Scenario 1 (light purple), Scenario 2 (orange), and Scenario 3 (red) are overlaid to show how parameter uncertainty is reduced as more information is incorporated.]
{Comparison of marginal posterior distributions across the three inference scenarios. 
Each panel shows the posterior for a given parameter (rows) and country (columns). 
The distributions for Scenario 1 (light purple), Scenario 2 (orange), and Scenario 3 (red) 
are overlaid to show how parameter uncertainty is reduced as more information is incorporated.}
\label{fig:appendix_posterior_comparison} \\

& \textbf{United States} 
& \textbf{Colombia} 
& \textbf{Dominican Rep.} 
& \textbf{Peru} \\[.6em]
\endfirsthead

\multicolumn{5}{c}{\figurename\ \thefigure{} -- continued from previous page} \\[.6em]
\endhead

\endfoot

& \textbf{United States} 
& \textbf{Colombia} 
& \textbf{Dominican Rep.} 
& \textbf{Peru} \\[.6em]
\endlastfoot

\textbf{$\mu_s$}     & \postplot{mu_s}{US}    & \postplot{mu_s}{CO}    & \postplot{mu_s}{DR}    & \postplot{mu_s}{PE} \\
\textbf{$\sigma_s$}  & \postplot{sigma_s}{US} & \postplot{sigma_s}{CO} & \postplot{sigma_s}{DR} & \postplot{sigma_s}{PE} \\
\textbf{$\delta_r$}  & \postplot{delta_r}{US} & \postplot{delta_r}{CO} & \postplot{delta_r}{DR} & \postplot{delta_r}{PE} \\
\textbf{$\sigma_r$}  & \postplot{sigma_r}{US} & \postplot{sigma_r}{CO} & \postplot{sigma_r}{DR} & \postplot{sigma_r}{PE} \\
\textbf{$\mu_d$}     & \postplot{mu_d}{US}    & \postplot{mu_d}{CO}    & \postplot{mu_d}{DR}    & \postplot{mu_d}{PE} \\
\textbf{$\sigma_d$}  & \postplot{sigma_d}{US} & \postplot{sigma_d}{CO} & \postplot{sigma_d}{DR} & \postplot{sigma_d}{PE} \\
\textbf{$\mu_b$}     & \postplot{mu_b}{US}    & \postplot{mu_b}{CO}    & \postplot{mu_b}{DR}    & \postplot{mu_b}{PE} \\
\textbf{$\sigma_b$}  & \postplot{sigma_b}{US} & \postplot{sigma_b}{CO} & \postplot{sigma_b}{DR} & \postplot{sigma_b}{PE} \\
\textbf{$\kappa$}    & \postplot{kappa}{US}   & \postplot{kappa}{CO}   & \postplot{kappa}{DR}   & \postplot{kappa}{PE} \\
\textbf{$\beta_{1}$} & \postplot{beta1}{US}   & \postplot{beta1}{CO}   & \postplot{beta1}{DR}   & \postplot{beta1}{PE} \\
\textbf{$\beta_{2}$} & \postplot{beta2}{US}   & \postplot{beta2}{CO}   & \postplot{beta2}{DR}   & \postplot{beta2}{PE} \\

\end{longtable}

\renewcommand\LTcaptype{table} 

\begin{figure}[H]
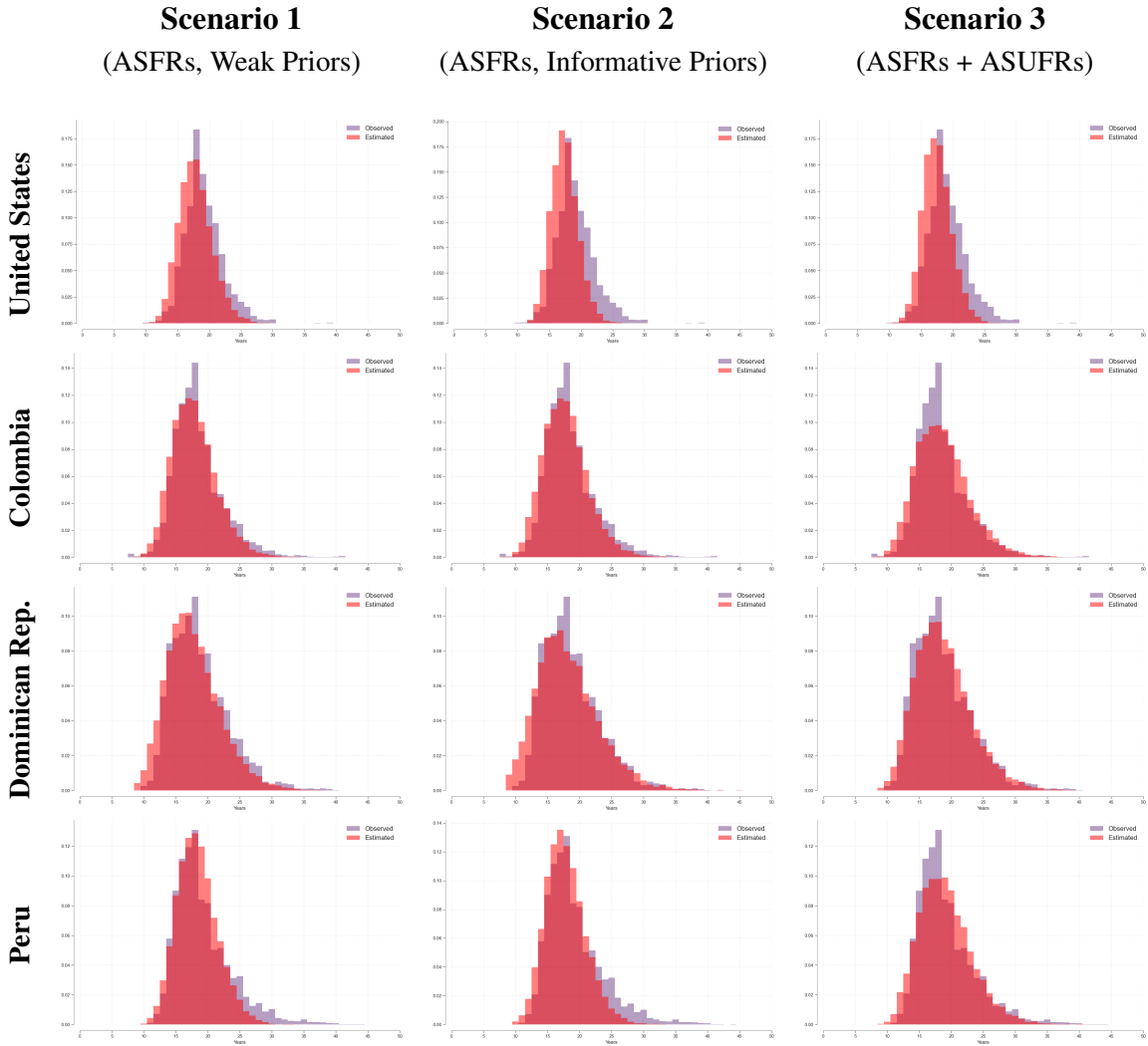
 
    \centering
    
    \newcommand{\valplot}[2]{%
        \includegraphics[width=0.3\linewidth]{plots/validation_age_start_sex_#1_s#2.png}%
    }

    \setlength{\tabcolsep}{4pt} 
    \begin{tabular}{lccc}
        & \textbf{Scenario 1} & \textbf{Scenario 2} & \textbf{Scenario 3} \\
        & \small(ASFRs, Weak Priors) & \small(ASFRs, Informative Priors) & \small(ASFRs + ASUFRs) \\[1em] 
        
        \raisebox{-.5\height}{\rotatebox{90}{\small\textbf{United States}}} &
        \valplot{US}{1} & \valplot{US}{2} & \valplot{US}{3} \\[1em]
        
        \raisebox{-.5\height}{\rotatebox{90}{\small\textbf{Colombia}}} &
        \valplot{CO}{1} & \valplot{CO}{2} & \valplot{CO}{3} \\[1em]

        \raisebox{-.05\height}{\rotatebox[origin=c]{90}{\small\textbf{Dominican Rep.}}} &
        \valplot{DR}{1} & \valplot{DR}{2} & \valplot{DR}{3} \\[1em]

        \raisebox{-.5\height}{\rotatebox{90}{\small\textbf{Peru}}} &
        \valplot{PE}{1} & \valplot{PE}{2} & \valplot{PE}{3} \\
    \end{tabular}

    \caption{Out-of-sample validation for the distribution of Age at First Sex. Each panel compares the observed survey distribution (purple bars) with the estimated posterior predictive distribution (red line) for each country (rows) and inference scenario (columns).}
    \label{fig:appendix_validation_afsex}
\end{figure}

Figure~\ref{fig:appendix_validation_afsex} presents the out-of-sample validation for the distribution of Age at First Sex across all four countries and three inference scenarios. Overall, the model demonstrates a strong ability to predict this key micro-level distribution, even under the baseline Scenario 1 which uses only aggregate ASFRs for estimation. In most cases, the model-implied posterior predictive distribution (red) closely approximates the shape and location of the observed survey data (purple).

Key aspects include a general smoothing of observed heaping (e.g., at age 18) by the model across scenarios, with initial mismatches like leftward shifts in the United States and overestimation of early debuts in Colombia and Peru in Scenario 1. For the Latin American countries, Scenario 3 shows clear enhancements, with red lines adhering more closely to the bars, better resolving tails and peaks. In contrast, the United States exhibits consistent predictions across all scenarios, with the red line remaining slightly left-shifted and showing no substantial improvement. 

\begin{figure}[H]
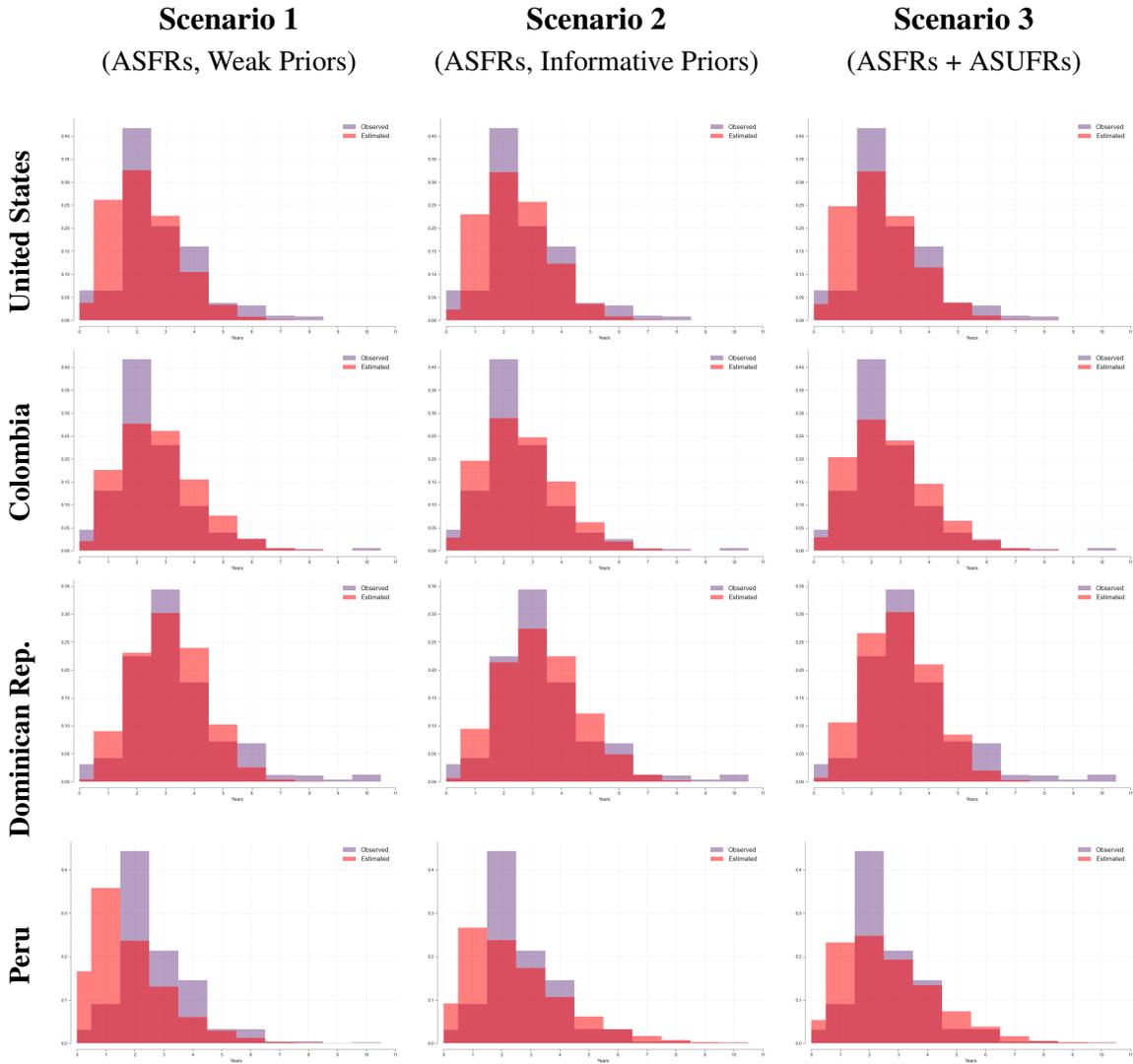
 
    \centering
    
    \newcommand{\valplotdfs}[2]{%
        \includegraphics[width=0.3\linewidth]{plots/validation_desired_family_size_#1_s#2.png}%
    }

    \setlength{\tabcolsep}{4pt}
    \begin{tabular}{lccc}
        & \textbf{Scenario 1} & \textbf{Scenario 2} & \textbf{Scenario 3} \\
        & \small(ASFRs, Weak Priors) & \small(ASFRs, Informative Priors) & \small(ASFRs + ASUFRs) \\[1em]
        
        \raisebox{-.5\height}{\rotatebox{90}{\small\textbf{United States}}} &
        \valplotdfs{US}{1} & \valplotdfs{US}{2} & \valplotdfs{US}{3} \\[1em]
        
        \raisebox{-.5\height}{\rotatebox{90}{\small\textbf{Colombia}}} &
        \valplotdfs{CO}{1} & \valplotdfs{CO}{2} & \valplotdfs{CO}{3} \\[1em]

        \raisebox{-.5\height}{\rotatebox[origin=c]{90}{\small\textbf{Dominican Rep.}}} &
        \valplotdfs{DR}{1} & \valplotdfs{DR}{2} & \valplotdfs{DR}{3} \\[1em]

        \raisebox{-.5\height}{\rotatebox{90}{\small\textbf{Peru}}} &
        \valplotdfs{PE}{1} & \valplotdfs{PE}{2} & \valplotdfs{PE}{3} \\
    \end{tabular}

    \caption{Out-of-sample validation for the distribution of Desired Family Size. Each panel compares the observed survey distribution (purple bars) with the model-implied posterior predictive distribution (red line) for each country (rows) and inference scenario (columns).}
    \label{fig:appendix_validation_dfs}
\end{figure}

Figure~\ref{fig:appendix_validation_dfs} shows the out-of-sample validation for Desired Family Size. Overall, the framework successfully recovers the general shape and rightward skew of the distribution in most cases. As noted in the main text, a consistent deviation is the model's treatment of the two-child norm; using a smooth parametric distribution, the model captures the central tendency of desired family size but smooths over the sharp empirical peak at exactly two children that results from a strong social norm.

The results for Peru, however, most clearly illustrate the value of incorporating additional information. In Scenario 1, the model struggles to correctly identify the mode of the distribution. The fit improves progressively from Scenario 2 to Scenario 3, where the predicted distribution aligns much more closely with the observed data. This highlights that, for some populations, incorporating external knowledge can be critical to achieve an accurate estimation of specific characteristic of the data.

\begin{figure}[H]
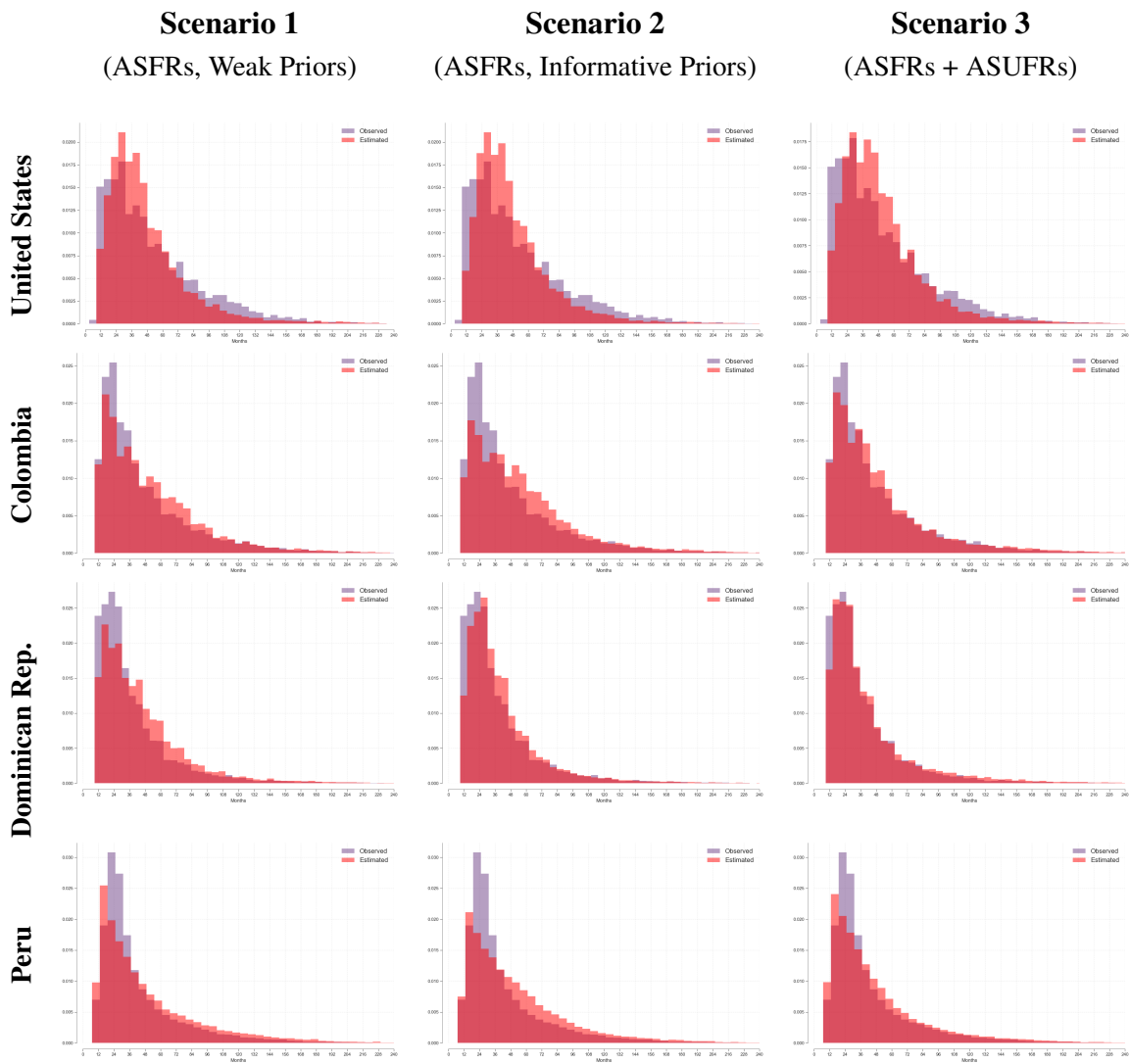
 
    \centering
    
    \newcommand{\valplotint}[2]{%
        \includegraphics[width=0.3\linewidth]{plots/validation_birth_intervals_#1_s#2.png}%
    }

    \setlength{\tabcolsep}{4pt}
    \begin{tabular}{lccc}
        & \textbf{Scenario 1} & \textbf{Scenario 2} & \textbf{Scenario 3} \\
        & \small(ASFRs, Weak Priors) & \small(ASFRs, Informative Priors) & \small(ASFRs + ASUFRs) \\[1em]
        
        \raisebox{-.5\height}{\rotatebox{90}{\small\textbf{United States}}} &
        \valplotint{US}{1} & \valplotint{US}{2} & \valplotint{US}{3} \\[1em]
        
        \raisebox{-.5\height}{\rotatebox{90}{\small\textbf{Colombia}}} &
        \valplotint{CO}{1} & \valplotint{CO}{2} & \valplotint{CO}{3} \\[1em]

        \raisebox{-.5\height}{\rotatebox[origin=c]{90}{\small\textbf{Dominican Rep.}}} &
        \valplotint{DR}{1} & \valplotint{DR}{2} & \valplotint{DR}{3} \\[1em]

        \raisebox{-.5\height}{\rotatebox{90}{\small\textbf{Peru}}} &
        \valplotint{PE}{1} & \valplotint{PE}{2} & \valplotint{PE}{3} \\
    \end{tabular}

    \caption{Out-of-sample validation for the distribution of Birth Intervals. Each panel compares the observed survey distribution (purple bars) with the model-implied posterior predictive distribution (red line) for each country (rows) and inference scenario (columns).}
    \label{fig:appendix_validation_intervals}
\end{figure}

Figure~\ref{fig:appendix_validation_intervals} presents the out-of-sample validation for the distribution of Birth Intervals. The model is broadly successful in recovering the principal spacing pattern across all countries, correctly identifying a peak in intervals around two to three years, followed by a long, declining tail.

The most notable discrepancy occurs at this peak. The model's predicted distribution (red line) often appears smoother and wider than the observed survey data (purple bars), which can exhibit a sharper peak at the most common spacing intervals. The value of incorporating additional information is most evident for the Dominican Republic. In Scenario 1, the location of the predicted peak is noticeably offset from the observed data, but it becomes progressively more accurate in Scenarios 2 and 3, eventually aligning almost perfectly with the empirical mode. For the other countries, the improvements across scenarios are more subtle, with the baseline model already providing a reasonable approximation of the distribution's central tendency, given model and data constraints.  

\bibliographystyle{chicago}
\bibliography{../../../../../../biblio/biblio}

\end{document}